\documentclass[aps,floatfix,amsmath,nofootinbib,amssymb,towcolumn,superscriptaddress]{revtex4-2}

\usepackage{overpic}
\usepackage{amssymb}
\usepackage{indentfirst}
\usepackage{feynmf}   
\usepackage{slashed}  
\usepackage{cases}
\usepackage{color}
\usepackage{multirow}
\usepackage{epstopdf}
\usepackage{graphicx,color,bm}
\usepackage{epstopdf}

\usepackage[colorlinks,
            citecolor=green,
            anchorcolor=red,
            menucolor=red,
            linkcolor=red,
            filecolor=red,
            runcolor=red,
            urlcolor=blue,
            frenchlinks=red]{hyperref}

\begin{document}

\title{Gluon GTMDs at nonzero skewness and impact parameter dependent parton distributions}

\author{Chentao Tan}\affiliation{School of Physics, Southeast University, Nanjing
211189, China}

\author{Zhun Lu}
\email{zhunlu@seu.edu.cn}
\affiliation{School of Physics, Southeast University, Nanjing 211189, China}

\begin{abstract}
We investigate the leading twist generalized transverse momentum dependent parton distributions (GTMDs) of the unpolarized and longitudinally polarized gluons in the nucleon. We adopt a light-front gluon-triquark model for the nucleon motivated by soft-wall AdS/QCD. The gluon GTMDs are defined through the off-forward gluon-gluon generalized correlator and are expressed as the overlap of light-cone wave functions. The GTMDs can be employed to provide the generalized parton distributions (GPDs) by integrating out the transverse momentum. The Fourier transform of the GPDs encodes the parton distributions in the transverse position space, namely, the impact parameter dependent parton distributions (IPDs). We also calculate the three gluon IPDs corresponding to the GPDs $H^g$, $E^g$ and $\widetilde{H}^g$, and present their dependence on $x$ and $b_\perp$, respectively.
\end{abstract}

\maketitle

\section{Introduction}\label{Sec:1}

An outstanding goal in hadron physics is to understand the structure of hadrons in terms of quarks and gluons. 
The deep inelastic scattering (DIS) is among the key tools to reveal hadronic structure, because one can extract the parton distribution functions (PDFs)~\cite{Collins:1981uw,Martin:1998sq,Gluck:1994uf,Gluck:1998xa} from such process. 
The PDFs are functions of the longitudinal momentum fraction, which encode the distributions of longitudinal momentum and polarizations of partons. 
A more comprehensive picture about the nucleon can be revealed by the transverse momentum dependent parton distributions (TMDs)~\cite{Mulders:2000sh,Meissner:2007rx}, which encode transverse motion of partons inside the nucleon. 
TMDs can be extracted from semi-inclusive reactions such as the semi-inclusive deep inelastic scatting (SIDIS) and the Drell-Yan process~\cite{Mulders:1995dh,Bacchetta:2006tn,Barone:2001sp,
Brodsky:2002cx,Bacchetta:2017gcc}. 
Besides the TMDs,  in the off-forward region a new type of nucleon structure--the so-call generalized parton distributions (GPDs)~\cite{Muller:1994ses,Ji:1996nm,Radyushkin:1997ki,Mueller:1998fv,
Goeke:2001tz,Diehl:2003ny,Ji:2004gf,
Belitsky:2005qn,Boffi:2007yc}--emerge.
It is the extension of the ordinary PDF from the forward scattering region to the off-forward scattering region.
The GPDs appear in the description of hard exclusive reactions, such as the deeply virtual Compton scattering(DVCS) and the deeply virtual meson production(DVMP)~\cite{Diehl:2003ny,Ji:1996nm,Belitsky:2005qn,Goeke:2001tz}.
Furthermore, the transverse position of partons is encoded in the impact parameter dependent parton distributions (IPDs)~\cite{Burkardt:2000za,Burkardt:2002hr}, which are the Fourier transform of the GPDs~\cite{Diehl:2003ny} at zero skewness with respect to the transverse momentum transfer of the hadron. 
 
The most complete structural information of hadrons is contained in the so-called generalized transverse momentum dependent parton distributions (GTMDs)~\cite{Meissner:2008ay,Meissner:2009ww}, which are often considered as the ``mother distributions", since several GTMDs can project to the TMDs and the GPDs in certain kinematical limits. 
The quark GTMDs may be measurable in the exclusive double Drell-Yan process~\cite{Bhattacharya:2017bvs}, while the feasibility to measure the gluon GTMDs in the diffractive dijet production has been studied~\cite{Hagiwara:2017fye,Ji:2016jgn,Hatta:2016aoc,Bhattacharya:2022vvo}.

The first complete classification of various parton distributions and their relations with each other has been discussed in Refs.~\cite{Meissner:2008ay,Meissner:2009ww}. 
There are sixteen twist-2 GTMDs for the quark and the gluon in the nucleon, respectively. These GTMDs encode the information of the distributions of the unpolarized and polarized partons. 
They are characterized by revealing the nucleon spin structure, for example, $F_{1,4}$ and $G_{1,1}$ play an important role in describing the canonical orbit angular momentum (OAM)~\cite{Lorce:2011kd,Hatta:2011ku,Ji:2012sj,Lorce:2012ce} and the spin-orbit correlations~\cite{Lorce:2014mxa,Tan:2021osk} of partons, respectively. The quark GTMDs for the nucleon have been calculated in various models, such as the light-cone constituent quark model~\cite{Lorce:2011ni,Lorce:2011kd,Lorce:2011dv}, the light-front dressed quark model~\cite{Mukherjee:2014nya,Mukherjee:2015aja,More:2017zqq}, the light-cone spectator model~\cite{Liu:2015eqa}, the light-front quark-diquark model~\cite{Chakrabarti:2016yuw,Chakrabarti:2017teq,Chakrabarti:2019wjx,Gutsche:2016gcd,Kaur:2019lox,Kumar:2017xcm}, the chiral soliton model~\cite{Lorce:2011ni,Lorce:2011kd} and the quark target model~\cite{Kanazawa:2014nha}, and so have these distributions for the pion~\cite{Ma:2018ysi,Kaur:2019jow,Kaur:2019kpi,Zhang:2021tnr}. 
In addition, the importance of the IPDs lies in their physical interpretation as a probability density in the impact parameter space. The quark IPDs for hadrons have been studied in Refs.~\cite{Broniowski:2003rp,Burkardt:2002hr}. 
However, the theoretical study on the gluon GTMDs and IPDs is still not sufficient.

It should be noted that most of the previous calculations for the GTMDs are made by assuming that the skewness $\xi$ is $0$. However, most of the data probed in the experiments is at $\xi \neq 0$. 
Therefore, a more detailed investigation of the GTMDs at nonzero skewness is necessary. 
In this work, we investigate the leading twist GTMDs of the unpolarized and longitudinally polarized gluons in the nucleon  using the soft-wall AdS/QCD model~\cite{Lyubovitskij:2021qza,Lyubovitskij:2020xqj}. 
This model has been widely applied to the calculations of the PDFs, form factors and mass spectrum~\cite{Brodsky:2014yha}. 
In the description of hadronic form factors at large $Q^2$~\cite{Brodsky:2014yha,Gutsche:2019jzh,Brodsky:2007hb,Abidin:2009hr}, the soft wall AdS/QCD embodies a main advantage, namely, the analytical implementation of quark counting rules~\cite{Brodsky:1973kr}. 
Same as the light-cone spectator model~\cite{Lu:2016vqu}, 
the minimum Fock state for the nucleon containing a gluon here is also a two-body composite system of a struck gluon and a three-quark spectator. 
We calculate the gluon GTMDs by writing the contracted correlators in the overlap representation in terms of the light-cone wave functions. 
In order to understand the gluon distributions in the transverse position space, we also calculate the three IPDs corresponding to the GPDs $H^g$, $E^g$ and $\widetilde{H}^g$.

The rest of the paper is organized as follows. 
In Sec.~\ref{Sec:2}, we provide the definitions of the gluon GTMDs and IPDs for the nucleon. 
In Sec.~\ref{Sec:3}, we write down the overlap representations of the contracted gluon correlators using the light-cone wave functions in the soft-wall AdS/QCD model. 
In Sec.~\ref{Sec:4}, we present not only the three-dimensional plots of the GTMDs, but also the dependence of the IPDs on $x$ and $b_\perp$, respectively. 
Summary of the paper is given in Sec.~\ref{Sec:5}.

\section{Definitions of gluon GTMDs at nonzero skewness and IPDs}\label{Sec:2}

A complete set of the gluon GTMDs for the nucleon has been presented in Ref.~\cite{Meissner:2009ww,Lorce:2013pza}. 
In the fixed light-cone time $z^+=0$, they can be defined through the off-forward gluon-gluon generalized correlator
\begin{align}
	W^{g[ij]}_{\lambda^\prime,\lambda}(x,P,\Delta,\bm{k}_\perp)=\int \frac{dz^-d^2\bm{z}_\perp}{(2\pi)^3 P^+}e^{ik\cdot z} \bigg\langle p^\prime,\lambda^\prime \bigg| F^{+i}_a(-\frac{z}{2}) \mathcal{W}_{ab}(-\frac{z}{2},\frac{z}{2}) F^{+j}_b(\frac{z}{2}) \bigg| p,\lambda \bigg\rangle \bigg|_{z^+=0}
	\label{eq:correlator},
\end{align}
where $p=P-\Delta/2$ ($p^\prime=P+\Delta/2$) and $\lambda$ ($\lambda^\prime$) represent the four-momentum and the helicity of the initial (final) nucleon, respectively. This correlator depends on four arguments. Among them,  $P=(p+p^\prime)/2$ and $\Delta=p^\prime-p$ denote the average four-momentum and the momentum transfer of the nucleon, respectively, $x=k^+/P^+$ denotes the average longitudinal gluon momentum fraction, and $\bm{k}_\perp$ denotes the average transverse gluon momentum. In Eq.~(\ref{eq:correlator}), the gluon field strength tensor $F_a^{\mu\nu}$ has a general form
\begin{align}
	F_a^{\mu\nu}=\partial^\mu A_a^\nu-\partial^\nu A_a^\mu + g f_{abc} A_b^\mu A_c^\nu,
\end{align}
where $f_{abc}$ and $g$ are the structure constants and the coupling constant of $\text{SU(3)}$, respectively. The Wilson line
\begin{align}
	\mathcal{W}_{ab}\bigg(-\frac{z}{2},\frac{z}{2}\bigg)\bigg|_{z^+=0}=
	\bigg[0^+,-\frac{z^-}{2},-\frac{\bm{z}_\perp}{2};0^+,\frac{z^-}{2},\frac{\bm{z}_\perp}{2}\bigg]_{ab}= \mathcal{P} \text{exp} \bigg[-g\int^\frac{z^-}{2}_{-\frac{z^-}{2}} dy^- \int^\frac{\bm{z}_\perp}{2}_{-\frac{\bm{z}_\perp}{2}} d\bm{y}_\perp f_{abc} A_c^+(0^+,y^-,\bm{y}_\perp)\bigg]
\end{align}
ensures the color gauge invariance of the bi-local gluon operator, where $\mathcal{P}$ denotes all possible ordered paths followed by the gluon field $A$. 
From now on, we choose the light-front gauge ($A^+=0$) and take the gauge link to be unit.

In leading twist, there are sixteen gluon GTMDs for the nucleon~\cite{Lorce:2013pza}. In this work, we concentrate on
\begin{align}
	W^g_{\lambda^\prime,\lambda}=&\delta_\perp^{ij} W^{g[ij]}_{\lambda^\prime,\lambda} \nonumber\\
	=&\frac{1}{2M}\bar{u}(p^\prime,\lambda^\prime)\bigg[F_{1,1}^g+\frac{i\sigma^{i+}k_\perp^i}{P^+}F^g_{1,2} +\frac{i\sigma^{i+}\Delta_\perp^i}{P^+}F^g_{1,3}+\frac{i\sigma^{ij}k_\perp^i\Delta^j_\perp}{M^2}F^g_{1,4}\bigg]u(p,\lambda) \nonumber\\
	=&\frac{1}{M\sqrt{1-\xi^2}} \bigg\{\bigg[M\delta_{\lambda^\prime,\lambda}-\frac{1}{2}(\lambda\Delta_\perp^1+i\Delta_\perp^2)\delta_{-\lambda^\prime,\lambda}\bigg]F^g_{1,1} +(1-\xi^2)(\lambda k_\perp^1+ik_\perp^2)\delta_{-\lambda^\prime,\lambda}F_{1,2}^g \nonumber \\
	&+(1-\xi^2)(\lambda\Delta_\perp^1+i\Delta_\perp^2)\delta_{-\lambda^\prime,\lambda}F^g_{1,3} +\frac{i\epsilon_\perp^{ij}k_\perp^i\Delta_\perp^j}{M^2} \bigg[\lambda M \delta_{\lambda^\prime,\lambda}-\frac{\xi}{2}(\Delta_\perp^1+i\lambda\Delta_\perp^2)\delta_{-\lambda^\prime,\lambda}\bigg]F_{1,4}^g \bigg\}
	\label{eq:F-GTMDs}
\end{align}
and
\begin{align}
	\widetilde{W}^g_{\lambda^\prime,\lambda}=&-i\epsilon_\perp^{ij} W^{g[ij]}_{\lambda^\prime,\lambda} \nonumber\\
	=&\frac{1}{2M}\bar{u}(p^\prime,\lambda^\prime)\bigg[-\frac{i\epsilon_\perp^{ij}k_\perp^i\Delta_\perp^j}{M^2} G_{1,1}^g+\frac{i\sigma^{i+}\gamma_5 k_\perp^i}{P^+}G^g_{1,2} +\frac{i\sigma^{i+}\gamma_5 \Delta_\perp^i}{P^+}G^g_{1,3}+i\sigma^{+-}\gamma_5 G^g_{1,4}\bigg]u(p,\lambda) \nonumber\\
	=&\frac{1}{M\sqrt{1-\xi^2}} \bigg\{-\frac{i\epsilon^{ij}_\perp k_\perp^i\Delta_\perp^j}{M^2} \bigg[M\delta_{\lambda^\prime,\lambda}-\frac{1}{2}(\lambda\Delta_\perp^1+i\Delta_\perp^2)\delta_{-\lambda^\prime,\lambda}\bigg]G^g_{1,1} +(1-\xi^2)(k_\perp^1+i\lambda k_\perp^2)\delta_{-\lambda^\prime,\lambda}G_{1,2}^g \nonumber \\
	&+(1-\xi^2)(\Delta_\perp^1+i\lambda\Delta_\perp^2)\delta_{-\lambda^\prime,\lambda}G^g_{1,3} +\bigg[\lambda M \delta_{\lambda^\prime,\lambda}-\frac{\xi}{2}(\Delta^1_\perp+i\lambda\Delta^2_\perp)\delta_{-\lambda^\prime,\lambda}\bigg]G_{1,4}^g \bigg\}
	\label{eq:G-GTMDs},
\end{align}
where $M$ denotes the nucleon mass, the skewness $\xi=(p^+-p^{\prime +})/(p^++p^{\prime +})=-\Delta^+/(2P^+)$ denotes the longitudinal momentum fraction transfer of the nucleon. In Eqs.~(\ref{eq:F-GTMDs},\ref{eq:G-GTMDs}), two tensors $\delta^{ij}_\perp=-g^{ij}_\perp$ and $\epsilon^{ij}_\perp=\epsilon^{+-ij}$ have been introduced, where $g^{\mu\nu}$ is the metric tensor and $\epsilon^{\alpha\beta\rho\sigma}$ is an antisymmetric tensor with $\epsilon^{+-12}=1$. Four F-type GTMDs describe the distributions of unpolarized gluons, while four G-type GTMDs describe the gluon helicity distributions. 
For transversely polarized gluons, one needs another set of eight leading twist H-type GTMDs. 
In general, the GTMDs are complex-valued functions~\cite{Meissner:2008ay,Meissner:2009ww}, and they depend on the variables $x$, $\xi$, $\bm{\Delta}_\perp^2$, $\bm{k}_\perp^2$ and $\bm{\Delta}_\perp \cdot \bm{k}_\perp$.
Then we express the gluon GTMDs in terms of the correlators contracted with two tensors as follows:
\begin{align} W^g_{+,+}(x,P,\Delta,\bm{k}_\perp)+W^g_{+,+}(x,P,\Delta,-\bm{k}_\perp)=&\frac{2}{M\sqrt{1-\xi^2}}MF_{1,1}^g
	\label{eq:f11},\\ W^g_{+,+}(x,P,\Delta,\bm{k}_\perp)-W^g_{+,+}(x,P,\Delta,-\bm{k}_\perp)=&\frac{2}{M\sqrt{1-\xi^2}} \frac{i\epsilon_\perp^{ij}k_\perp^i \Delta_\perp^j}{M^2} M F_{1,4}^g
	\label{eq:f14},\\ W^g_{-,+}(x,P,\Delta,\bm{k}_\perp)+W^g_{-,+}(x,P,\Delta,-\bm{k}_\perp)=&\frac{2}{M\sqrt{1-\xi^2}} \bigg\{ -\frac{1}{2} (\Delta_\perp^1+i\Delta_\perp^2)F_{1,1}^g+(1-\xi^2)(\Delta_\perp^1+i\Delta_\perp^2)F_{1,3}^g \bigg\}
	\label{eq:f13},\\	W^g_{-,+}(x,P,\Delta,\bm{k}_\perp)-W^g_{-,+}(x,P,\Delta,-\bm{k}_\perp)=&\frac{2}{M\sqrt{1-\xi^2}} \bigg\{(1-\xi^2) (k_\perp^1+ik_\perp^2)F_{1,2}^g \nonumber\\ &+\frac{i\epsilon_\perp^{ij}k_\perp^i\Delta_\perp^j}{M^2}\bigg[-\frac{\xi}{2}(\Delta_\perp^1+i\Delta_\perp^2)\bigg]F_{1,4}^g\bigg\}
	\label{eq:f12},
\end{align}
for F-type GTMDs, and
\begin{align}	\widetilde{W}^g_{+,+}(x,P,\Delta,\bm{k}_\perp)+\widetilde{W}^g_{+,+}(x,P,\Delta,-\bm{k}_\perp)=&\frac{2}{M\sqrt{1-\xi^2}}MG_{1,4}^g
	\label{eq:g14},\\ \widetilde{W}^g_{+,+}(x,P,\Delta,\bm{k}_\perp)-\widetilde{W}^g_{+,+}(x,P,\Delta,-\bm{k}_\perp)=&\frac{2}{M\sqrt{1-\xi^2}} \bigg[-\frac{i\epsilon_\perp^{ij}k_\perp^i \Delta_\perp^j}{M^2}M G_{1,1}^g\bigg]
	\label{eq:g11},\\ \widetilde{W}^g_{-,+}(x,P,\Delta,\bm{k}_\perp)+\widetilde{W}^g_{-,+}(x,P,\Delta,-\bm{k}_\perp)=&\frac{2}{M\sqrt{1-\xi^2}} \bigg\{ (1-\xi^2)(\Delta_\perp^1+i\Delta_\perp^2)G_{1,3}^g -\frac{\xi}{2}(\Delta_\perp^1+i\Delta_\perp^2)G_{1,4}^g\bigg\}
	\label{eq:g13},\\	\widetilde{W}^g_{-,+}(x,P,\Delta,\bm{k}_\perp)-\widetilde{W}^g_{-,+}(x,P,\Delta,-\bm{k}_\perp)=&\frac{2}{M\sqrt{1-\xi^2}} \bigg\{ - \frac{i\epsilon_\perp^{ij} k_\perp^i \Delta_\perp^j}{M^2} \bigg[-\frac{1}{2} (\Delta_\perp^1+i\Delta_\perp^2)\bigg] G_{1,1}^g \nonumber\\
	&+(1-\xi^2)(k_\perp^1+ik_\perp^2)G_{1,2}^g \bigg\}
	\label{eq:g12},
\end{align}
for G-type GTMDs,
where the subscript $+$ ($-$) denotes that the helicity of the nucleon is $+1/2$ ($-1/2$).

There are four leading twist gluon GPDs that can be expressed in terms of the GTMDs in Eqs.~(\ref{eq:F-GTMDs},\ref{eq:G-GTMDs}) as~\cite{Maji:2022tog}
\begin{align}
	H^g(x,\xi,\bm{\Delta}^2_\perp)=&\int d^2\bm{k}_\perp \bigg[ F_{1,1}^g+2\xi^2\bigg( \frac{\bm{\Delta}_\perp \cdot \bm{k}_\perp }{\bm{\Delta}^2_\perp}F_{1,2}^g+F_{1,3}^g \bigg) \bigg]
	\label{Hg},\\
	E^g(x,\xi,\bm{\Delta}^2_\perp)=&\int d^2\bm{k}_\perp \bigg[-F_{1,1}^g+2(1-\xi^2)\bigg(\frac{\bm{\Delta}_\perp \cdot \bm{k}_\perp }{\bm{\Delta}^2_\perp}F_{1,2}^g+F_{1,3}^g\bigg)\bigg]
	\label{Eg},\\
	\widetilde{H}^g(x,\xi,\bm{\Delta}^2_\perp)=&\int d^2\bm{k}_\perp \bigg[ 2\xi \bigg( \frac{\bm{\Delta}_\perp \cdot \bm{k}_\perp }{\bm{\Delta}^2_\perp}G_{1,2}^g+G_{1,3}^g \bigg)+G_{1,4}^g \bigg]
	\label{Hgtilde},\\
	\widetilde{E}^g(x,\xi,\bm{\Delta}^2_\perp)=&\int d^2\bm{k}_\perp \bigg[\frac{2(1-\xi^2)}{\xi} \bigg(\frac{\bm{\Delta}_\perp \cdot \bm{k}_\perp }{\bm{\Delta}^2_\perp}G_{1,2}^g+G_{1,3}^g \bigg) -G_{1,4}^g \bigg]
	\label{Egtilde}.
\end{align}

The IPDs are defined as the Fourier transform of the GPDs at zero skewness with respect to $\bm{\Delta}_\perp$. 
At $\xi=0$, $\widetilde{E}^g$ will not show up, and the other three GPDs provide the corresponding IPDs $\mathcal{H}^g$, $\mathcal{E}^g$ and $\widetilde{\mathcal{H}}^g$ through the Fourier transform
\begin{align}
	\mathcal{X}(x,\bm{b}_\perp)=&\int \frac{d^2 \bm{\Delta}_\perp}{(2\pi)^2}e^{-i\bm{\Delta}_\perp \cdot \bm{b}_\perp} X(x,0,\bm{\Delta}^2_\perp),
\end{align}
where $\bm{b}_\perp$ is a two-dimensional vector in the impact parameter space conjugate to $\bm{\Delta}_\perp$. 
The IPD $\mathcal{H}^g$ is related to the transverse position distribution of unpolarized gluons in an unpolarized nucleon, $\mathcal{E}^g$ is related to the transverse position distribution of unpolarized gluons in a transversely polarized nucleon, and $\widetilde{\mathcal{H}}^g$ is related to the transverse position distribution of longitudinally polarized gluons in a longitudinally polarized nucleon.

\section{gluon correlators in soft-wall AdS/QCD model}\label{Sec:3}

In this section, we express the contracted gluon correlators as the overlap of the light-cone wave functions in the soft-wall AdS/QCD model. 
The light-cone formalism has been widely used in different theoretical approaches to reveal the structural information of hadrons~\cite{More:2017zqq,Bacchetta:2008af,Brodsky:2000xy}. On the other hand, the soft-wall AdS/QCD model has also been successfully employed to describe the structural information of nucleons, e.g., various parton distributions, electromagnetic form factors, spin asymmetries, etc.,~\cite{Lyubovitskij:2020xqj,Maji:2022tog}. The authors in Ref.~\cite{Maji:2022tog} presented the quark GTMDs at nonzero skewness for the nucleon in this model, then the investigation of these distributions for gluons will be equally important. Similarly, $|qqqg\rangle$ is considered as the minimum Fock state for a nucleon containing a gluon. We treat this four-body system as a two-body composite system of a spin-1 gluon $g$ and a spin-1/2 three-quark spectator $X$ for simplicity~\cite{Lu:2016vqu}:
\begin{align}
	|p;S \rangle \rightarrow |g_{s_g}X_{s_X}(qqq)\rangle,
\end{align}
where $s_g$ and $s_X$ denote the gluon and spectator spins, respectively. We introduce the notation of the light-cone wave function $\psi^{\lambda}_{\lambda_g \lambda_X}(x,\bm{k}_\perp)$, where $\lambda$, $\lambda_g$ and $\lambda_X$ denote the helicity of the nucleon, the gluon and the three-quark spectator, respectively. 
Then the Fock-state expansion for a nucleon with helicity $\lambda=+1/2$ (we use the shorten notation $+$) is given by
\begin{align}
	|\Psi^+_{\text{two \, particle}}(p^+,\bm{p}_\perp=\bm{0}_\perp) \rangle =&\int \frac{d^2\bm{k}_\perp dx}{16\pi^3\sqrt{x(1-x)}}\nonumber \\
	&\times \left[\psi^+_{+1+\frac{1}{2}}(x,\bm{k}_\perp)\bigg|+1,+\frac{1}{2};xp^+,\bm{k}_\perp \bigg\rangle+\psi^+_{+1-\frac{1}{2}}(x,\bm{k}_\perp)\bigg|+1,-\frac{1}{2};xp^+,\bm{k}_\perp \bigg\rangle \nonumber \right.\\
	&\left.+\psi^+_{-1+\frac{1}{2}}(x,\bm{k}_\perp)\bigg|-1,+\frac{1}{2};xp^+,\bm{k}_\perp \bigg\rangle+\psi^+_{-1-\frac{1}{2}}(x,\bm{k}_\perp)\bigg|-1,-\frac{1}{2};xp^+,\bm{k}_\perp \bigg\rangle \right],
\end{align}
where
\begin{align}	\psi^+_{+1+\frac{1}{2}}(x,\bm{k}_\perp)&=\frac{k^1-ik^2}{M}\phi^{(2)}(x,\bm{k}_\perp^2),\nonumber\\
	\psi^+_{+1-\frac{1}{2}}(x,\bm{k}_\perp)&=\phi^{(1)}(x,\bm{k}_\perp^2),\nonumber\\
	\psi^+_{-1+\frac{1}{2}}(x,\bm{k}_\perp)&=-\frac{k^1+ik^2}{M}(1-x)\phi^{(2)}(x,\bm{k}_\perp^2),\nonumber\\
	\psi^+_{-1-\frac{1}{2}}(x,\bm{k}_\perp)&=0.
	\label{eq:wavefunction+}
\end{align}

The Fock-state expansion for a nucleon with helicity $\lambda=-1/2$ (we use the shorten notation $-$) has the form
\begin{align}
	|\Psi^-_{\text{two \, particle}}(p^+,\bm{p}_\perp=\bm{0}_\perp) \rangle =&\int \frac{d^2\bm{k}_\perp dx}{16\pi^3\sqrt{x(1-x)}}\nonumber \\
	&\times \left[\psi^-_{+1+\frac{1}{2}}(x,\bm{k}_\perp)\bigg|+1,+\frac{1}{2};xp^+,\bm{k}_\perp \bigg\rangle+\psi^-_{+1-\frac{1}{2}}(x,\bm{k}_\perp)\bigg|+1,-\frac{1}{2};xp^+,\bm{k}_\perp \bigg\rangle \nonumber \right.\\
	&\left.+\psi^-_{-1+\frac{1}{2}}(x,\bm{k}_\perp)\bigg|-1,+\frac{1}{2};xp^+,\bm{k}_\perp \bigg\rangle+\psi^-_{-1-\frac{1}{2}}(x,\bm{k}_\perp)\bigg|-1,-\frac{1}{2};xp^+,\bm{k}_\perp \bigg\rangle \right],
\end{align}
where
\begin{align}
	\psi^-_{+1+\frac{1}{2}}(x,\bm{k}_\perp)&=0,\nonumber\\
	\psi^-_{+1-\frac{1}{2}}(x,\bm{k}_\perp)&=-[\psi^+_{-1+\frac{1}{2}}(x,\bm{k}_\perp)]^* \nonumber\\
	&=\frac{k^1-ik^2}{M}(1-x)\phi^{(2)}(x,\bm{k}_\perp^2),\nonumber\\
	\psi^-_{-1+\frac{1}{2}}(x,\bm{k}_\perp)&=+[\psi^+_{+1-\frac{1}{2}}(x,\bm{k}_\perp)]^* \nonumber\\
	&=\phi^{(1)}(x,\bm{k}_\perp^2),\nonumber\\
	\psi^-_{-1-\frac{1}{2}}(x,\bm{k}_\perp)&=-[\psi^+_{+1+\frac{1}{2}}(x,\bm{k}_\perp)]^* \nonumber\\
	&=-\frac{k^1+ik^2}{M}\phi^{(2)}(x,\bm{k}_\perp^2).
	\label{eq:wavefunction-}
\end{align}

The functions $\phi^{(1)}(x,\bm{k}_\perp^2)$ and $\phi^{(2)}(x,\bm{k}_\perp^2)$ in Eqs.~(\ref{eq:wavefunction+},\ref{eq:wavefunction-}) can be expressed in terms of the gluon PDFs
\begin{align}
	G^{\pm}(x)=&\frac{G(x)\pm \Delta G(x)}{2},
\end{align}
where
\begin{align}
	G(x)=&f^g_1(x)=\int d^2 \bm{k}_\perp f_1^g(x,\bm{k}_\perp^2), \nonumber \\
	\Delta G(x)=&g^g_{1L}(x)=\int d^2 \bm{k}_\perp g_{1L}^g(x,\bm{k}_\perp^2),
\end{align}
as
\begin{align}	\phi^{(1)}(x,\bm{k}_\perp^2)=&\frac{4\pi}{\kappa}\sqrt{G^+(x)}\beta(x)\sqrt{D_g(x)}\text{exp}\bigg[-\frac{\bm{k}_\perp^2}{2\kappa^2}D_g(x)\bigg], \nonumber \\
	\frac{1}{M}\phi^{(2)}(x,\bm{k}_\perp^2)=&\frac{4\pi}{\kappa^2}\sqrt{G^-(x)} \frac{D_g(x)}{1-x}\text{exp}\bigg[-\frac{\bm{k}_\perp^2}{2\kappa^2}D_g(x)\bigg],
	\label{eq:phi12}
\end{align}
where
\begin{align}
	\beta(x)=&\sqrt{1-\frac{G^-(x)}{G^+(x)(1-x)^2}}.
\end{align}

The parameter $\kappa$ is related to the nucleon mass, as $M=2\sqrt{2}\kappa$~\cite{Gutsche:2011vb,Gutsche:2019jzh}. The profile function $D_g(x)$ can be connected to the profile function $f_g(x)$ as~\cite{Lyubovitskij:2020otz}
\begin{align}
	D_g(x)=&-\frac{\text{log}[1-[f_g(x)]^{2/5}(1-x)^2]}{(1-x)^2},
\end{align}
where $f_g(x)$ is the solution of the following differential equation with respect to $x$:
\begin{align}
	[-f_g(x)(1-x)^5]^\prime=xG(x).
\end{align}

Based on the above light-cone wave functions, the overlap representations of the contracted correlators in Eqs.~(\ref{eq:f11}-\ref{eq:g12}) can be expressed as
\begin{align} W^g_{\lambda,\lambda}(x,P,\Delta,\bm{k}_\perp)&=N\sum_{\lambda_g,\lambda_X}\psi^{\lambda\star}_{\lambda_g,\lambda_X}(x^{\text{out}},\bm{k}_\perp^{\text{out}}) \psi^{\lambda}_{\lambda_g,\lambda_X}(x^{\text{in}},\bm{k}_\perp^{\text{in}})(\epsilon_{\lambda_g}^1\epsilon_{\lambda_g}^{1\ast}+\epsilon_{\lambda_g}^2\epsilon_{\lambda_g}^{2\ast})
	\label{eq:w1+},\\ W^g_{-,+}(x,P,\Delta,\bm{k}_\perp)&=-N\sum_{\lambda_g,\lambda_X}\psi^{-\star}_{\lambda_g,\lambda_X}(x^{\text{out}},\bm{k}_\perp^{\text{out}}) \psi^{+}_{\lambda_g,\lambda_X}(x^{\text{in}},\bm{k}_\perp^{\text{in}})(\epsilon_{\lambda_g}^1\epsilon_{\lambda_g}^{1\ast}+\epsilon_{\lambda_g}^2\epsilon_{\lambda_g}^{2\ast})
	\label{eq:w1-},\\ \widetilde{W}^g_{\lambda,\lambda}(x,P,\Delta,\bm{k}_\perp)&=iN\sum_{\lambda_g,\lambda_X}\psi^{\lambda\star}_{\lambda_g,\lambda_X}(x^{\text{out}},\bm{k}_\perp^{\text{out}}) \psi^{\lambda}_{\lambda_g,\lambda_X}(x^{\text{in}},\bm{k}_\perp^{\text{in}})(\epsilon_{\lambda_g}^1\epsilon_{\lambda_g}^{2\ast}-\epsilon_{\lambda_g}^2\epsilon_{\lambda_g}^{1\ast})
	\label{eq:w2+},\\ \widetilde{W}^g_{-,+}(x,P,\Delta,\bm{k}_\perp)&=-iN\sum_{\lambda_g,\lambda_X}\psi^{-\star}_{\lambda_g,\lambda_X}(x^{\text{out}},\bm{k}_\perp^{\text{out}}) \psi^{+}_{\lambda_g,\lambda_X}(x^{\text{in}},\bm{k}_\perp^{\text{in}})(\epsilon_{\lambda_g}^1\epsilon_{\lambda_g}^{2\ast}-\epsilon_{\lambda_g}^2\epsilon_{\lambda_g}^{1\ast})
	\label{eq:w2-},
\end{align}
where $N=1/(2(2\pi)^3)$. The gluon polarization vectors $\epsilon_{\lambda_g}^\mu$ read
\begin{align}
	\epsilon^\mu_{\pm}=(0,0,\bm{\epsilon}_{\pm})=\frac{1}{\sqrt{2}}(0,0,\mp1,-i).
\end{align}

The initial and final transverse momenta of the struck gluon are given by
\begin{align}
	\bm{k}^{\text{in}}_\perp=&\bm{k}_\perp - (1-x^{\text{in}})\frac{\bm{\Delta}_\perp}{2},\qquad \text{with} \qquad x^{\text{in}}=\frac{x+\xi}{1+\xi}, \\
	\bm{k}^{\text{out}}_\perp=&\bm{k}_\perp +(1-x^{\text{out}}) \frac{\bm{\Delta}_\perp}{2},\qquad \text{with} \qquad x^{\text{out}}=\frac{x-\xi}{1-\xi},
	\label{eq:xin}
\end{align}
respectively.

\section{Numerical results}\label{Sec:4}
\begin{figure}
	\centering
	\includegraphics[width=0.4\columnwidth]{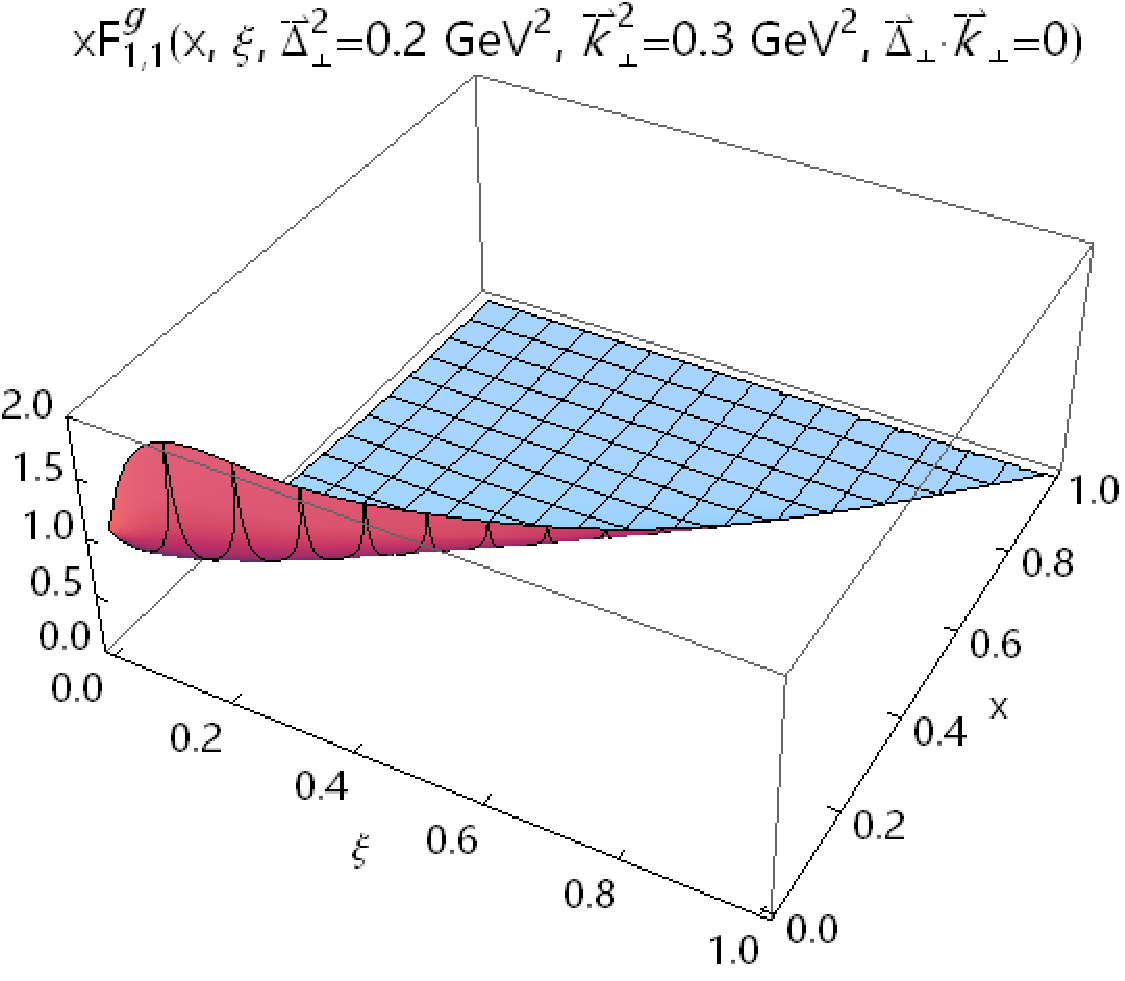}~~~
	\includegraphics[width=0.4\columnwidth]{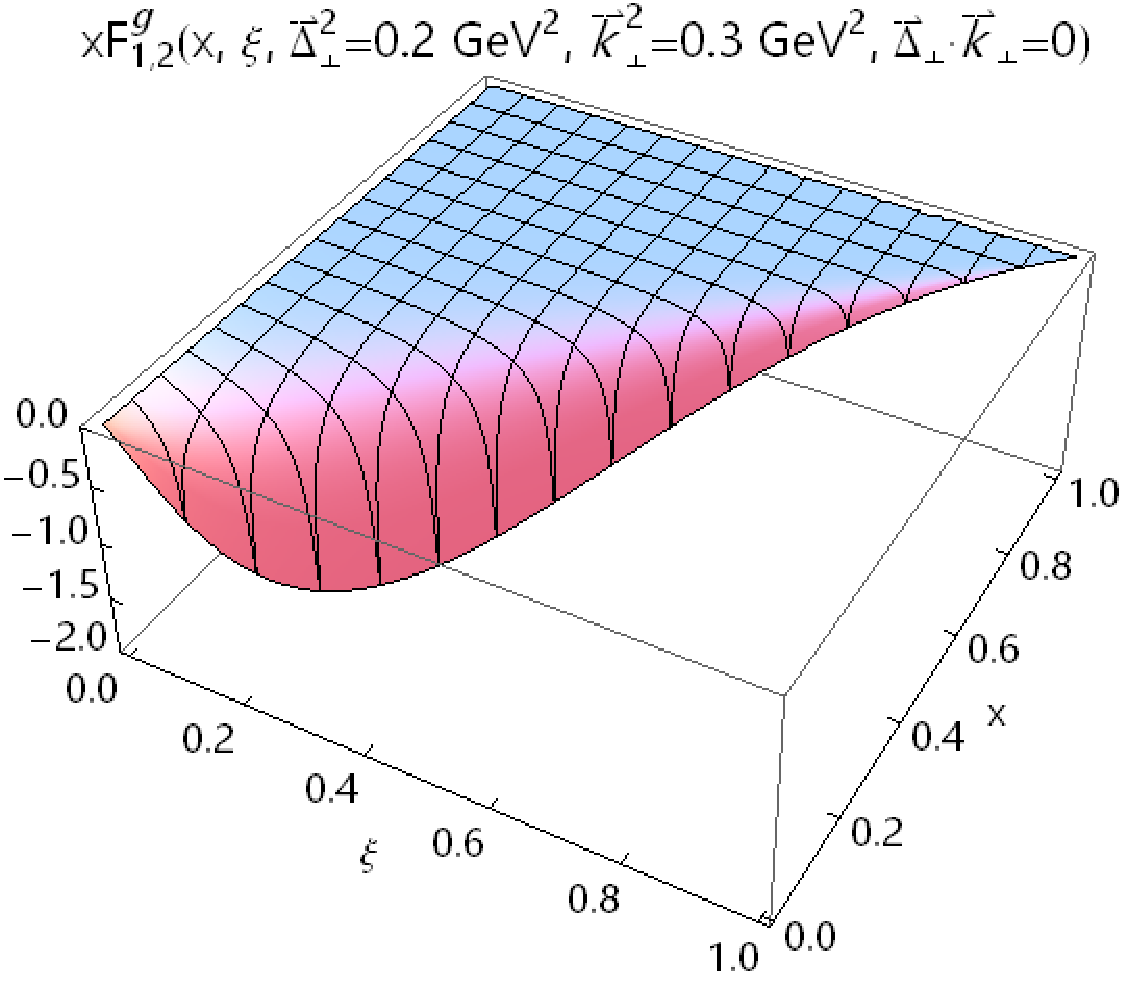}\\
	\includegraphics[width=0.4\columnwidth]{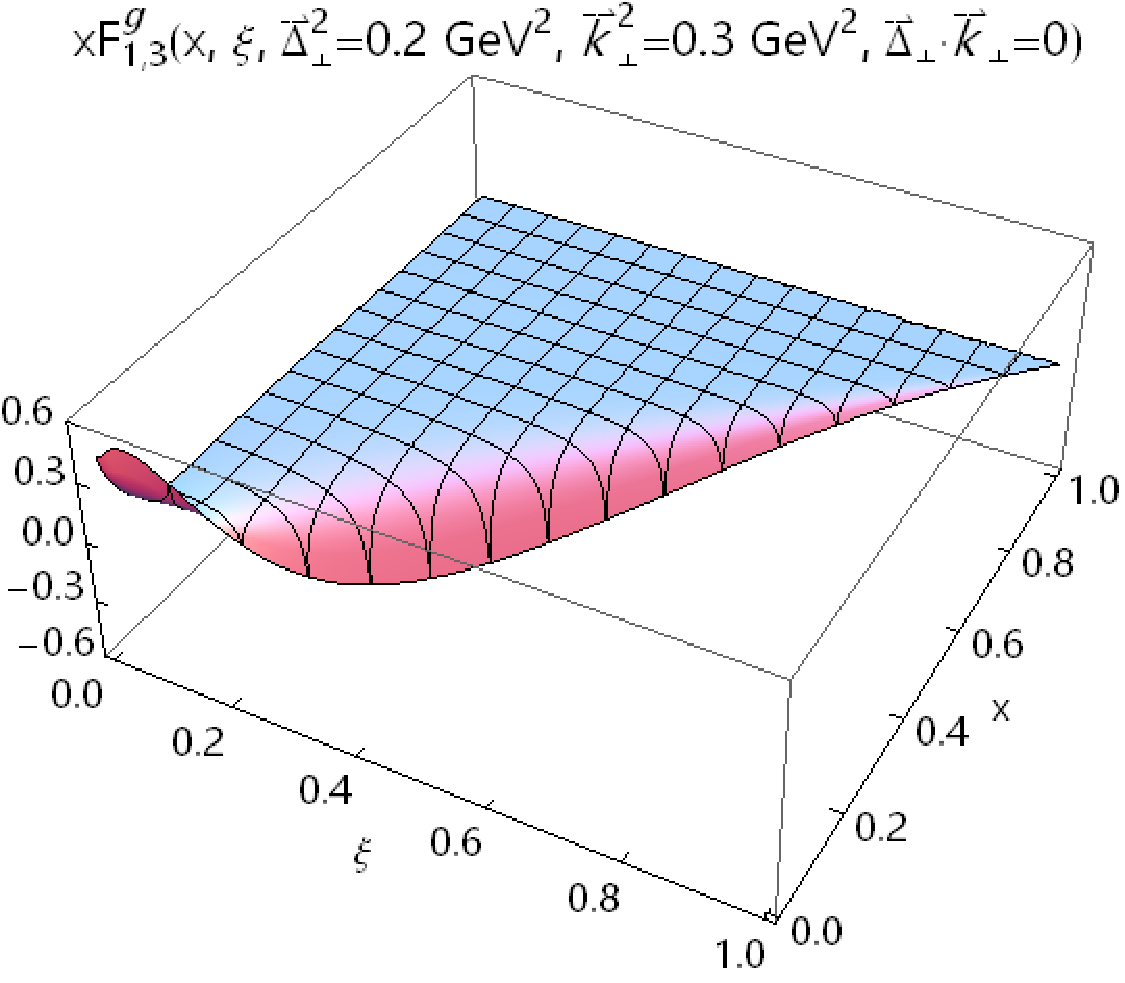}~~~
	\includegraphics[width=0.4\columnwidth]{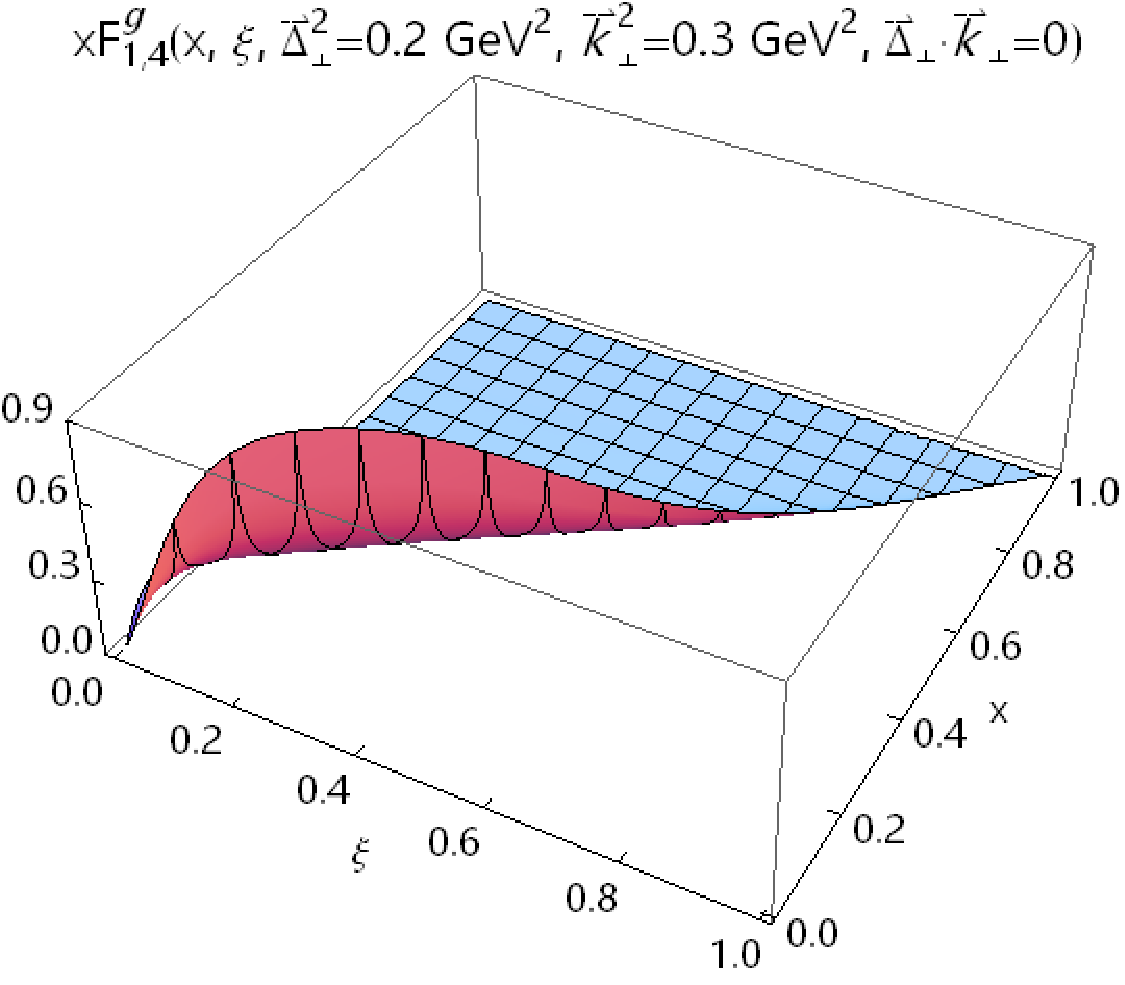}\\
	\caption{The GTMDs (timed with $x$) as functions of $x$ and $\xi$ for an unpolarized gluon at fixed $\bm{\Delta}^2_\perp=0.2 \ \text{GeV}^2$ and $\bm{k}_\perp^2=0.3 \ \text{GeV}^2$.}
	\label{fig:fxxi}
\end{figure}

In this section, we present the numerical results of the gluon GTMDs and IPDs. By substituting the light-cone wave functions into the overlap representation of the correlators, we get the analytical results for the gluon GTMDs in terms of the unpolarized gluon PDF $G(x)$ and the longitudinally polarized gluon PDF $\Delta G(x)$, which can be taken from world data analysis. In Ref.~\cite{Sufian:2020wcv}, the authors provide three parametrizations for the gluon PDFs based on the QCD formalism developed in Refs.~\cite{Brodsky:1989db,Brodsky:1994kg} and the NNPDF global analysis~\cite{NNPDF:2017mvq,Nocera:2014gqa}. They are almost equivalent and we choose the ``ansatz-3" parametrization, which reads~\cite{Sufian:2020wcv}
\begin{align}
	xG^+(x)=&x^\alpha[A(1-x)^{4+\beta}+B(1-x)^{5+\beta}](1+\gamma\sqrt{x}+\delta x), \nonumber\\
	xG^-(x)=&x^\alpha[A(1-x)^{6+\beta^\prime}+B(1-x)^{7+\beta^\prime}](1+\gamma^\prime \sqrt{x}+\delta^\prime x),
\end{align}
where the parameters are fixed as
\begin{align}
	\alpha&=0.034, \quad \beta=0.54, \quad \gamma=-2.63, \quad \delta=2.54,\nonumber\\
	\beta^\prime&=2.00, \qquad \gamma^\prime=-2.80, \quad \delta^\prime=2.25.
	\label{eq:parameter}
\end{align}
The normalization parameters $A$ and $B$ are determined from the moments of the gluon PDFs $\langle x_g \rangle$ and $\Delta G$, where the former (the second-$x$ moment of $G(x)$) represents the total gluon momentum fraction and the latter (the first-$x$ moment of $\Delta G (x)$) represents the total gluon helicity. We take $A=14.09$ and $B=-11.24$ by globally analysing the unpolarized gluon PDF in Ref.~\cite{Sufian:2020wcv}. Finally, in view of the good fit between the model results and the experimental data in Ref.~\cite{Lyubovitskij:2020xqj}, the parameter $\kappa$ is fixed as 0.383 GeV.

\subsection{GTMDs for an unpolarized gluon}
\begin{figure}
	\centering
	\includegraphics[width=0.4\columnwidth]{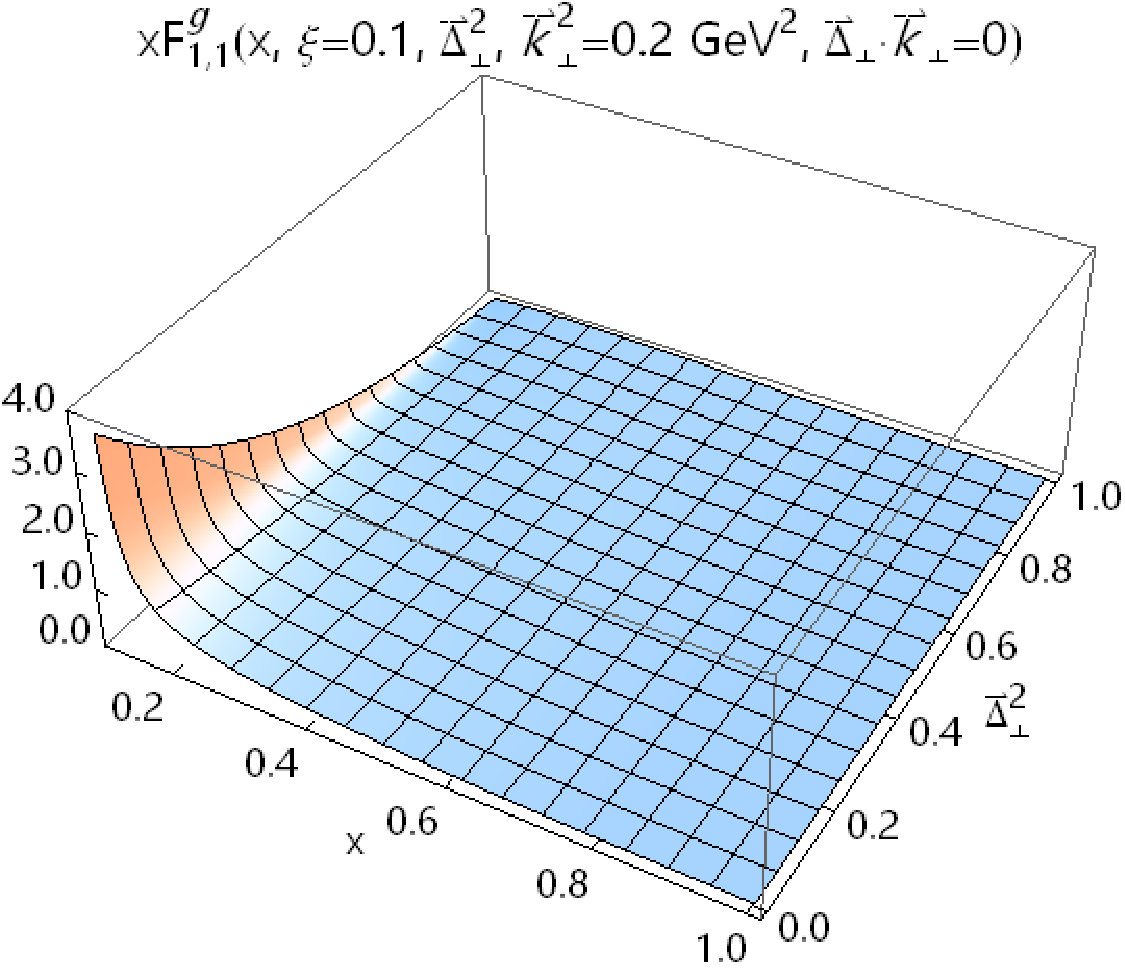}~~~
	\includegraphics[width=0.4\columnwidth]{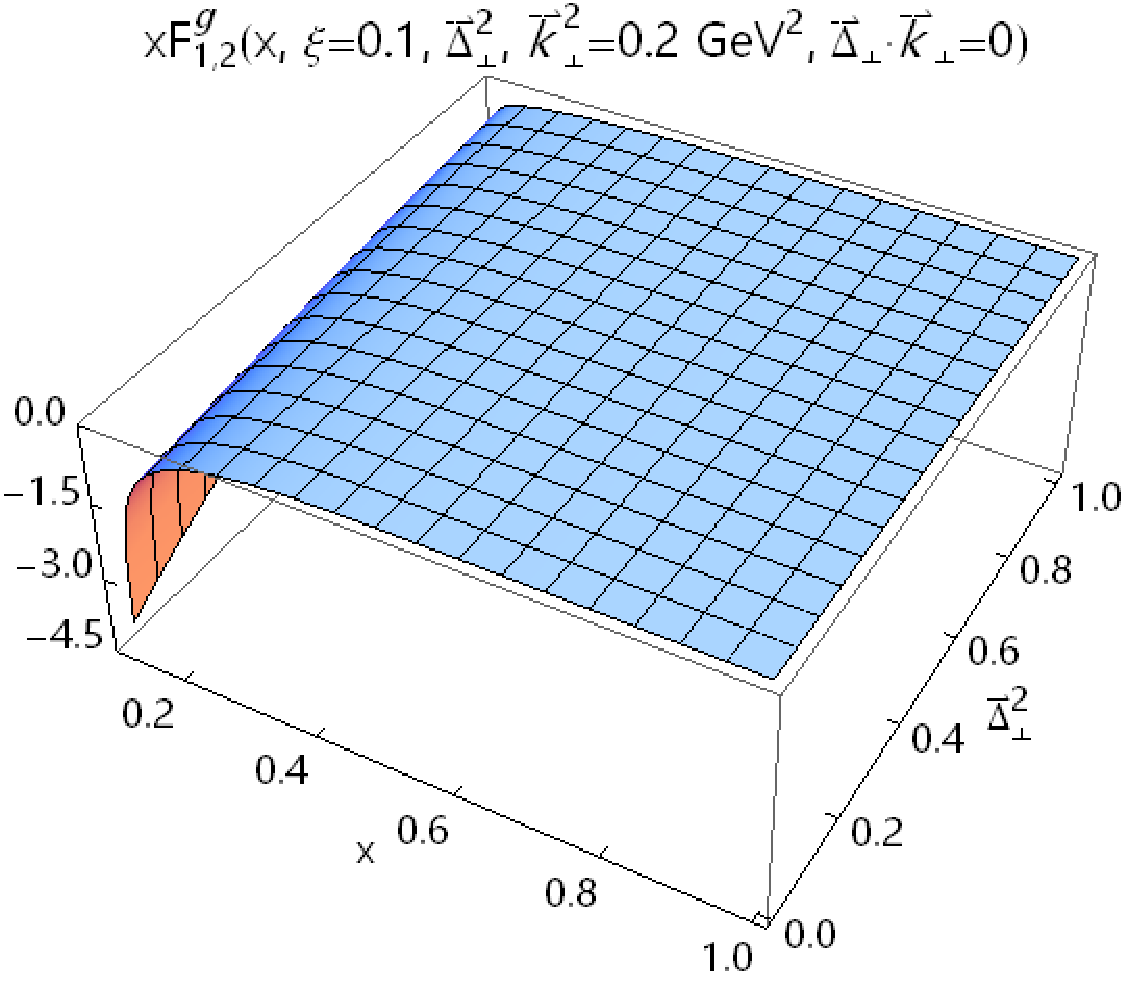}\\
	\includegraphics[width=0.4\columnwidth]{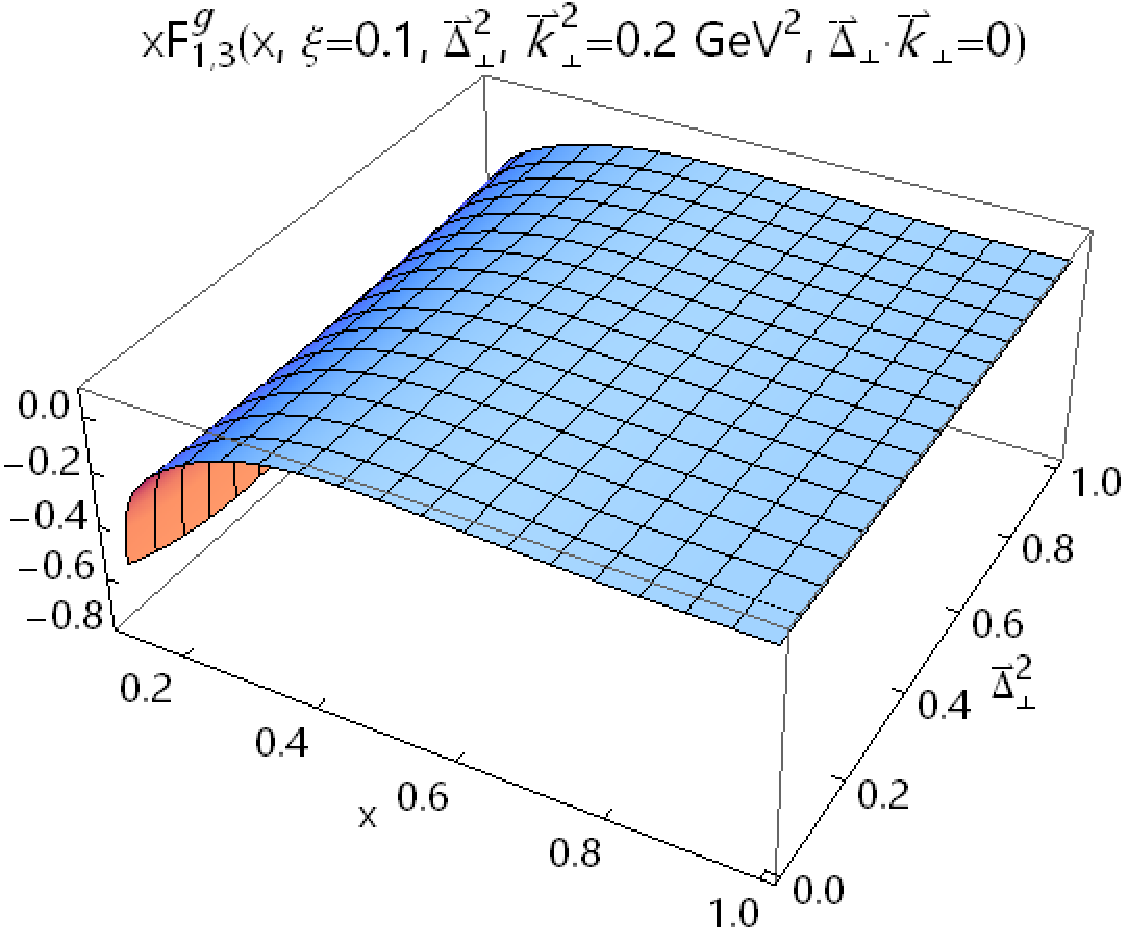}~~~
	\includegraphics[width=0.4\columnwidth]{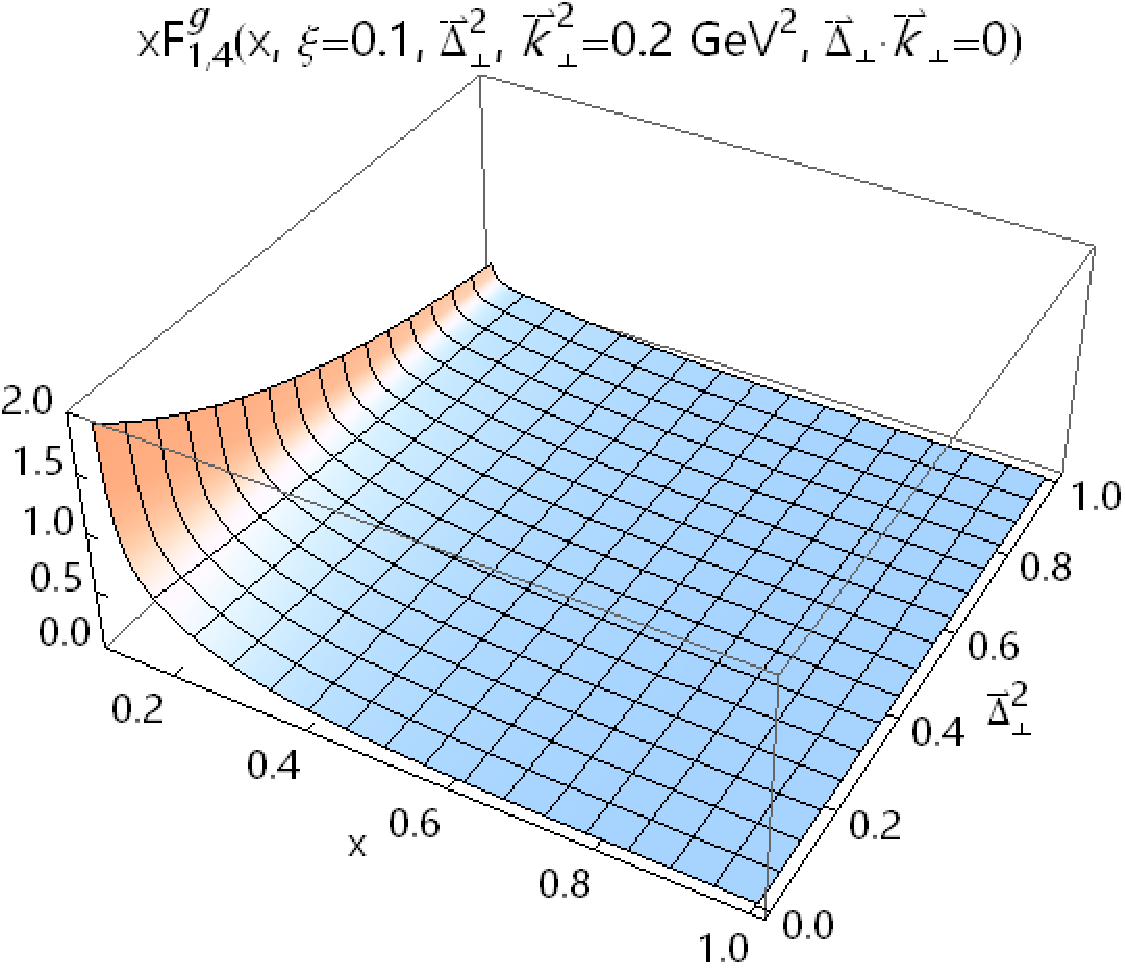}\\
	\caption{The GTMDs (timed with $x$) as functions of $x$ and $\bm{\Delta}_\perp^2$ for an unpolarized gluon at fixed $\xi=0.1$ and $\bm{k}_\perp^2=0.2 \ \text{GeV}^2$.}
	\label{fig:fxde}
\end{figure}

In Fig.~\ref{fig:fxxi}, we present the dependence of the GTMDs (timed with $x$) for an unpolarized gluon in a nucleon on $x$ and $\xi$ at fixed $\bm{\Delta}^2_\perp=0.2 \ \text{GeV}^2$ and $\bm{k}_\perp^2=0.3 \ \text{GeV}^2$ with $\bm{\Delta}_\perp \perp \bm{k}_\perp$. 
We observe that all the distributions exhibit the accessibility in the DGLAP region $x>\xi$, where $xF^g_{1,1}$ and $xF^g_{1,4}$ show the positive distributions, $xF^g_{1,2}$ shows the negative distribution, while $xF_{1,3}$ is positive in the lower-$x$($<0.1$) region and is negative in the higher-$x$ region. 
Referring to the definition of $F_{1,3}$ in Eq.~(\ref{eq:f13}), we find that $F_{1,1}^g$ dominates at low-$x$, which leads to the positive distribution of $xF^g_{1,3}$ at $x<0.1$. As the longitudinal momentum transfer to the nucleon $\Delta^+$ increases, the peaks of the GTMDs (timed with $x$), which are always at the limit of $x \rightarrow \xi$, shift toward the higher values of $x$, and the magnitudes decrease from the low-$x$ until they disappear. 
Our model results are obtained in the light-cone gauge, so the canonical gluon OAM $l_z^g$ can be defined through $F_{1,4}^g$ at $\xi=0$ and $\bm{\Delta}_\perp=0$ as
\begin{align}
	l_z^g=-\int dx d^2\bm{k}_\perp \frac{\bm{k}_\perp^2}{M^2}F_{1,4}^g(x,0,0,\bm{k}_\perp^2,0),
\end{align}
which gives the intrinsic gluon OAM, independent from spectator interactions, and encodes the correlation between the gluon OAM and the nucleon spin. The positive distribution of $F_{1,4}^g$ indicates that $l_z^g<0$, that is, the total gluon OAM will reduce the contribution of the gluon angular momentum to the nucleon spin. 
On the other hand, this result also means that the gluon OAM and the nucleon spin tend to be antialigned, which is consistent with the result of the light-cone spectator model~\cite{Tan:2023vvi}.

Fig.~\ref{fig:fxde} shows the dependence of the unpolarized gluon GTMDs (timed with $x$) on $x$ and $\bm{\Delta}_\perp^2$ at fixed $\xi=0.1$ and $\bm{k}_\perp^2=0.2 \ \text{GeV}^2$ with $\bm{\Delta}_\perp \perp \bm{k}_\perp$. We note that the overall shape of all the plots is similar. The peaks of these distributions are always at the limit of $x \rightarrow 0$ and the magnitudes decrease with increasing $\bm{\Delta}_\perp^2$. 
On the other hand, with the increase of the light-cone fraction $x$, which indicates that the gluon carries a larger portion of the nucleon momentum, the magnitudes also decrease, but at a much higher rate. 
Therefore, the main contributions of the gluon GTMDs are concentrated at low-$x$, and this is also a common property of the gluon PDFs, TMDs and GPDs~\cite{Meissner:2007rx,Lyubovitskij:2020xqj}. 
In the TMD limit ($\Delta=0$), the T-even part of $F_{1,1}^g$ projects to the unpolarized gluon TMD $f_1^g(x,\bm{k}_\perp^2)$, and the T-odd part of $F_{1,2}^g$ is linked to the gluon Sivers function $-f_{1T}^{\perp g}(x,\bm{k}_\perp^2)$~\cite{Meissner:2009ww,Bhattacharya:2018lgm}. 
In the GPD limit (integrating over $\bm{k}_\perp$), the projection from the GTMDs $F_{1,1}^g$, $F_{1,2}^g$ and $F_{1,3}^g$ to the GPDs $H^g$ and $E^g$ have been shown in Eqs.~(\ref{Hg},\ref{Eg}).

\subsection{GTMDs for a longitudinally polarized gluon}
\begin{figure}
	\centering
	\includegraphics[width=0.4\columnwidth]{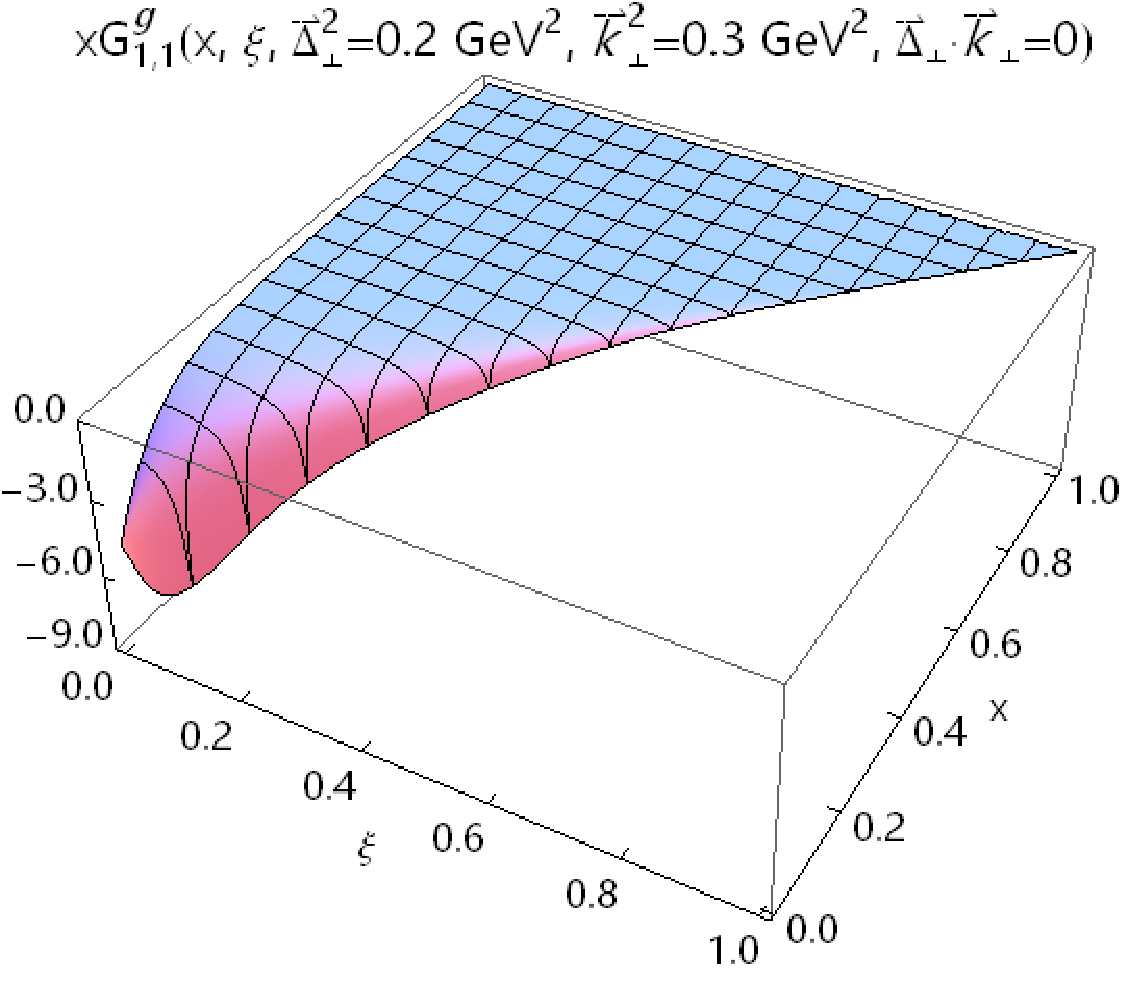}~~~
	\includegraphics[width=0.4\columnwidth]{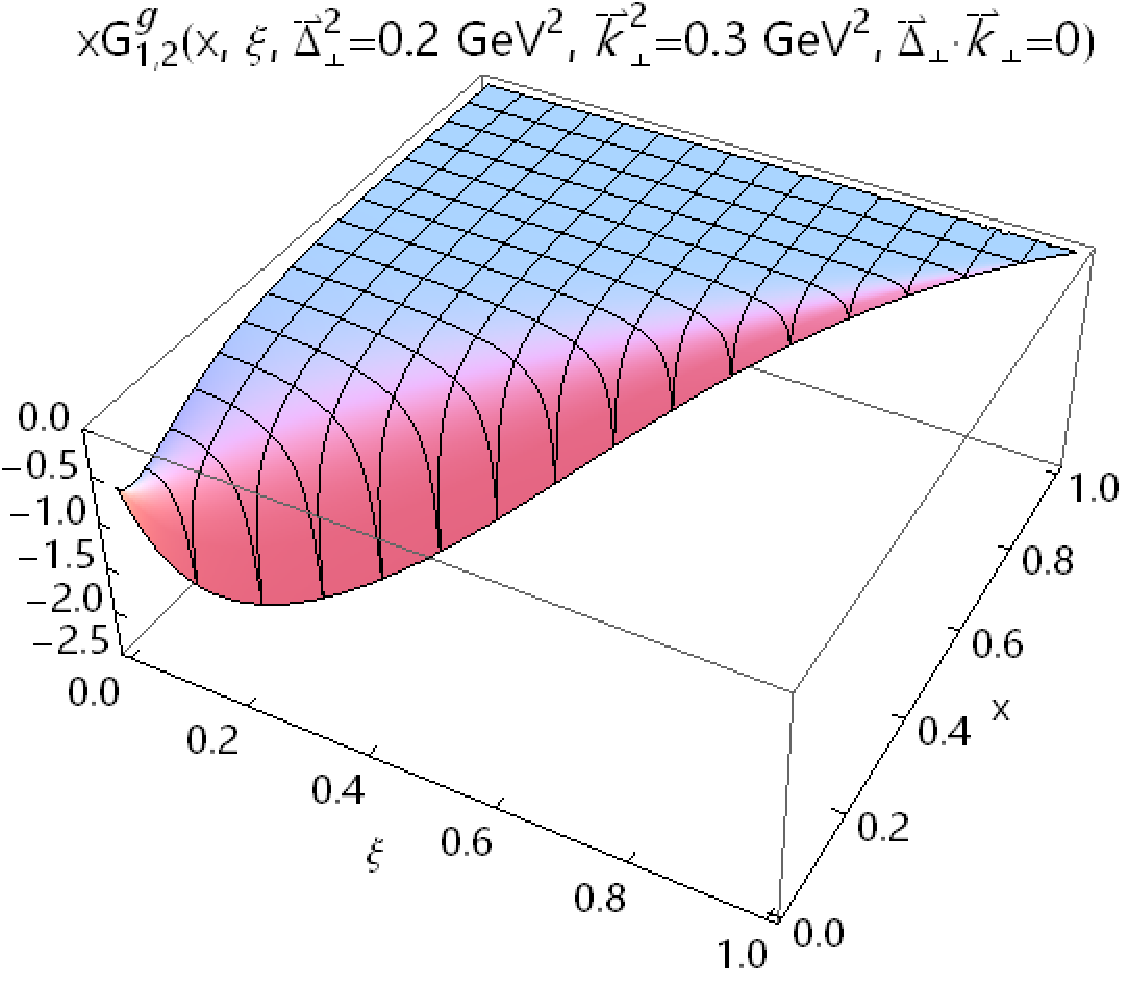}\\
	\includegraphics[width=0.4\columnwidth]{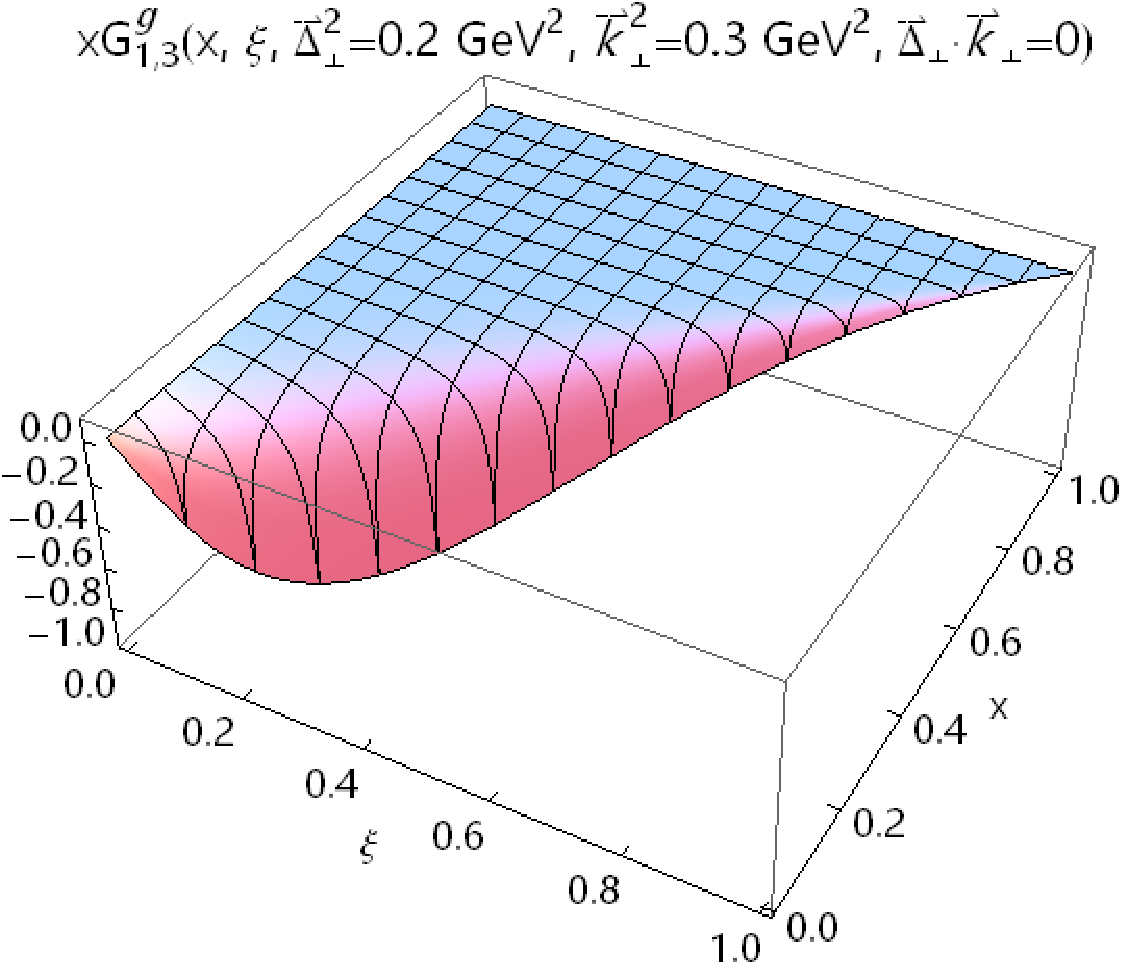}~~~
	\includegraphics[width=0.4\columnwidth]{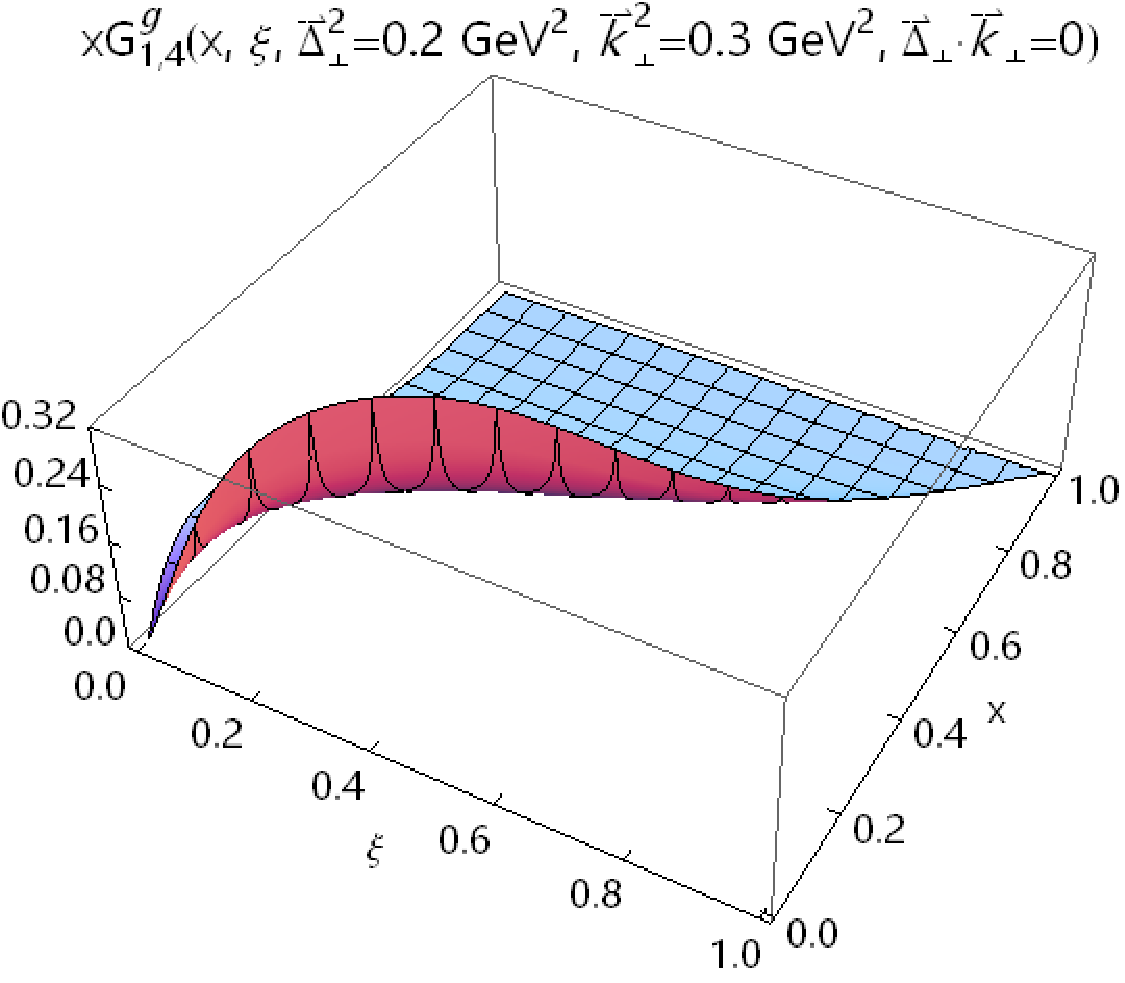}\\
	\caption{The GTMDs (timed with $x$) as functions of $x$ and $\xi$ for a longitudinally polarized gluon at fixed $\bm{\Delta}^2_\perp=0.2 \ \text{GeV}^2$ and $\bm{k}_\perp^2=0.3 \ \text{GeV}^2$.}
	\label{fig:gxxi}
\end{figure}

In Fig.~\ref{fig:gxxi}, we present the dependence of the GTMDs (timed with $x$) for a longitudinally polarized gluon in a nucleon on $x$ and $\xi$ at fixed $\bm{\Delta}^2_\perp=0.2 \ \text{GeV}^2$ and $\bm{k}_\perp^2=0.3 \ \text{GeV}^2$ with $\bm{\Delta}_\perp \perp \bm{k}_\perp$. As mentioned earlier, all the distributions are only accessible in the DGLAP region $x>\xi$. 
We observe that $xG_{1,1}^g$ and $xG_{1,2}^g$ are negative, $xG_{1,4}^g$ is positive, while $xG_{1,3}^g$ is negative at lower-$x$($<0.5$) and it is slightly positive at higher-$x$. 
Referring to the definition of $G_{1,3}$ in Eq.~(\ref{eq:g13}), we find that the contribution of $G_{1,4}^g$ is largely suppressed in the whole $x>\xi$ region. 
Moreover, the qualitative behaviors of the peaks and magnitudes of these distributions with increasing $\Delta^+$ are the same as those of the unpolarized GTMDs. 
Similarly, the canonical gluon spin-orbit correlations $C_z^g$ can be defined through $G_{1,1}^g$ at $\Delta=0$ as
\begin{align}
	C_z^g=\int dx d^2\bm{k}_\perp \frac{\bm{k}_\perp^2}{M^2}G_{1,1}^g(x,0,0,\bm{k}_\perp^2,0).
\end{align}
The negative $G_{1,1}^g$ implies that the gluon spin and OAM tend to be antialigned in our model, which is also consistent with the result of the light-cone spectator model~\cite{Tan:2023vvi}. 
In the TMD limit, the T-even part of $G_{1,4}^g$ reduces to the gluon helicity TMD $g_{1L}^g(x,\bm{k}_\perp^2)$, and the T-even part of $G_{1,2}^g$ reduces to the gluon worm-gear TMD $g_{1T}^g(x,\bm{k}_\perp^2)$~\cite{Meissner:2009ww,Bhattacharya:2018lgm}. 
In the GPD limit, the projection from the GTMDs $G_{1,2}^g$, $G_{1,3}^g$ and $G_{1,4}^g$ to the GPDs $\tilde{H}^g$ and $\tilde{E}^g$ have been shown in Eqs.~(\ref{Hgtilde},\ref{Egtilde}).

\begin{figure}
	\centering
	\includegraphics[width=0.4\columnwidth]{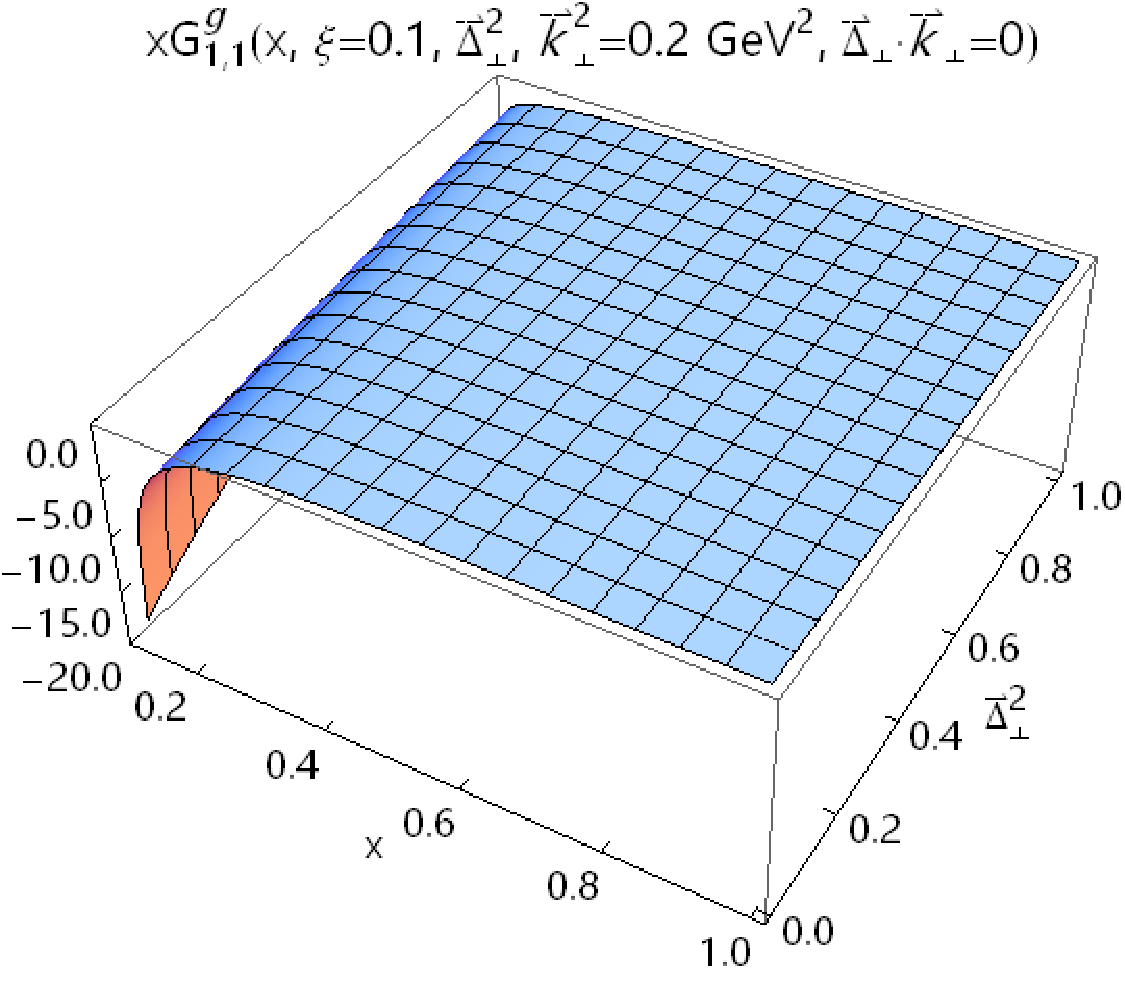}~~~
	\includegraphics[width=0.4\columnwidth]{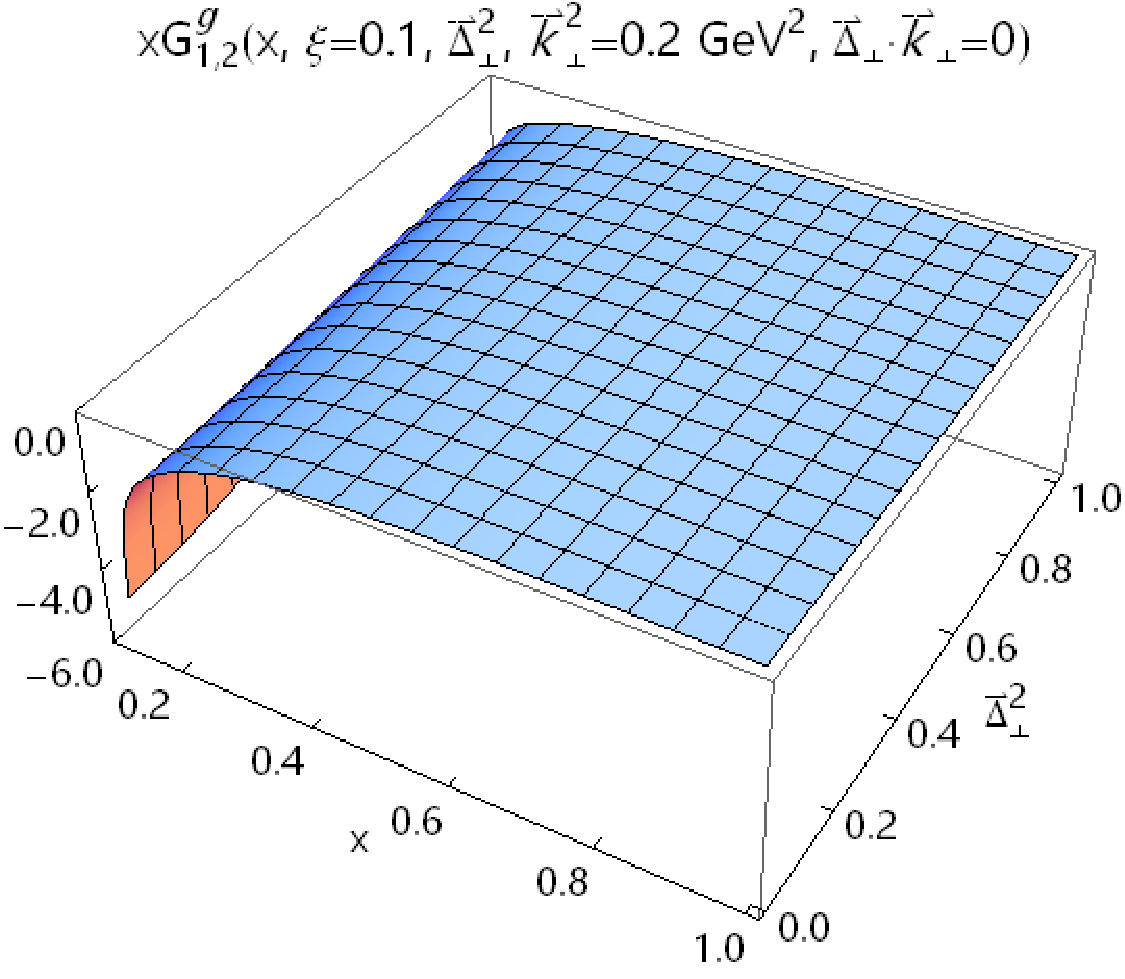}\\
	\includegraphics[width=0.4\columnwidth]{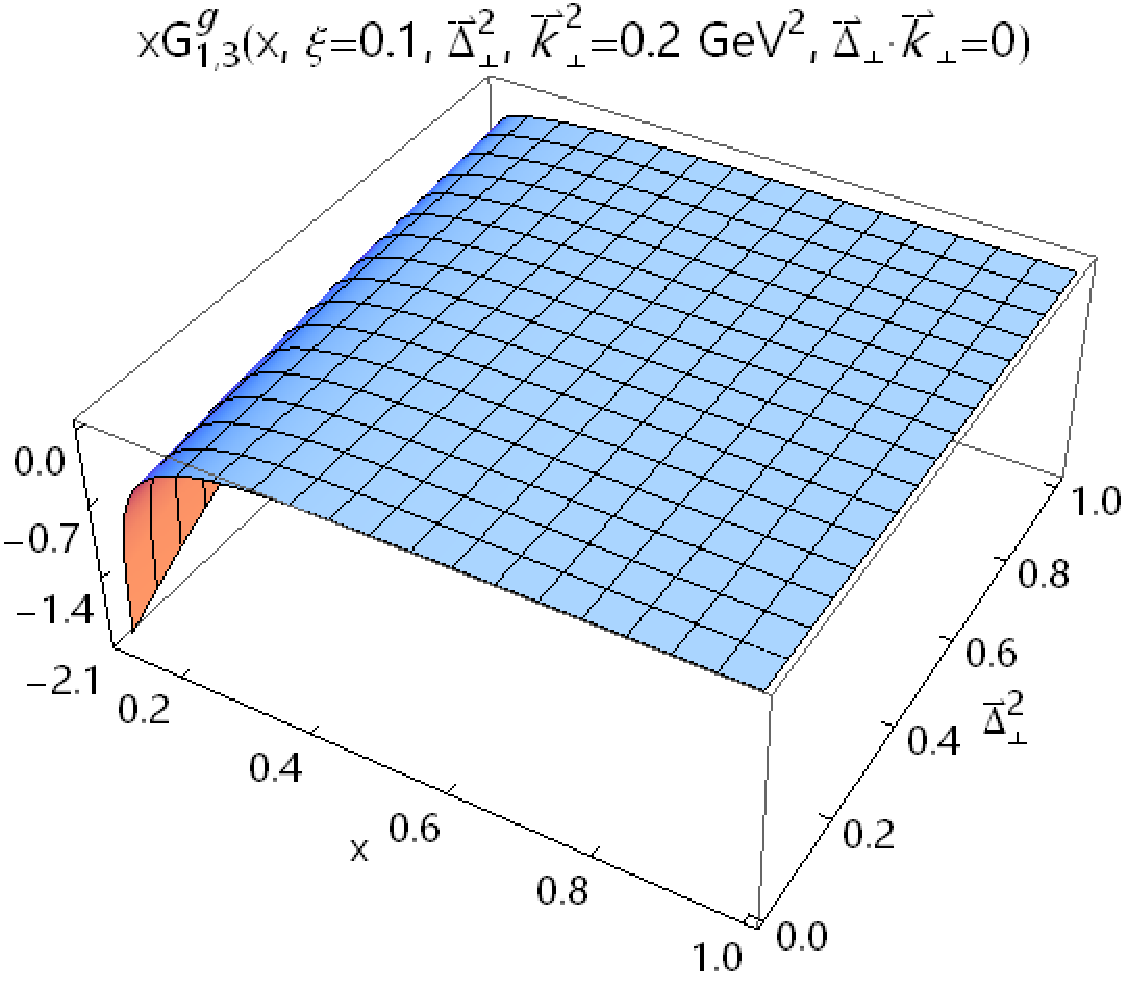}~~~
	\includegraphics[width=0.4\columnwidth]{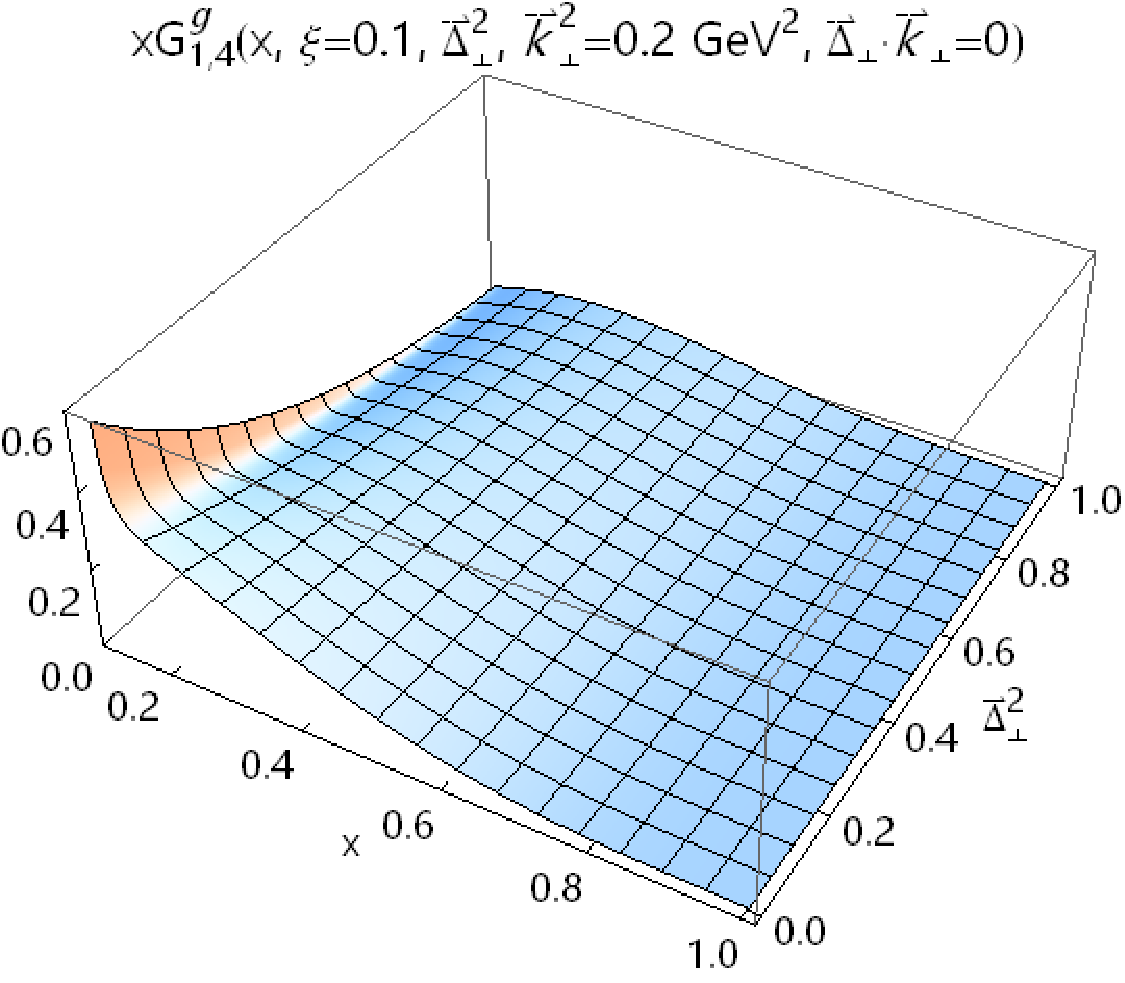}\\
	\caption{The GTMDs (timed with $x$) as functions of $x$ and $\bm{\Delta}_\perp^2$ for a longitudinally polarized gluon at fixed $\xi=0.1$ and $\bm{k}_\perp^2=0.2 \ \text{GeV}^2$.}
	\label{fig:gxde}
\end{figure}

In Fig.~\ref{fig:gxde}, we present the dependence of the longitudinally polarized gluon GTMDs (timed with $x$) on $x$ and $\bm{\Delta}_\perp^2$ at fixed $\xi=0.1$ and $\bm{k}_\perp^2=0.2 \ \text{GeV}^2$ with $\bm{\Delta}_\perp \perp \bm{k}_\perp$. 
We find that the qualitative properties of these distributions are similar to those of the unpolarized GTMDs.

\subsection{IPDs $\mathcal{H}^g$, $\mathcal{E}^g$ and $\widetilde{\mathcal{H}}^g$}
\begin{figure}
	\centering
	\includegraphics[width=0.42\columnwidth]{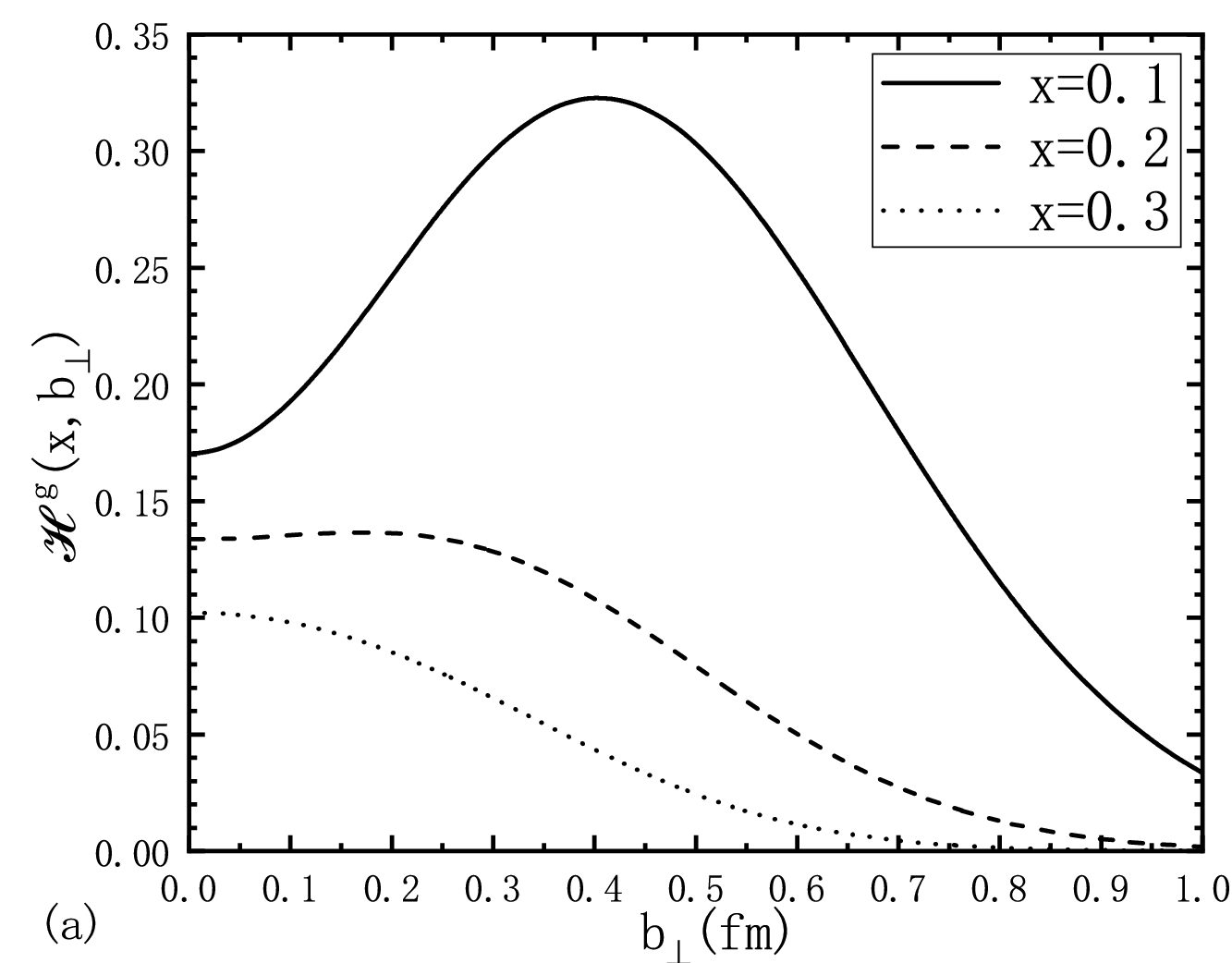}~~~
	\includegraphics[width=0.42\columnwidth]{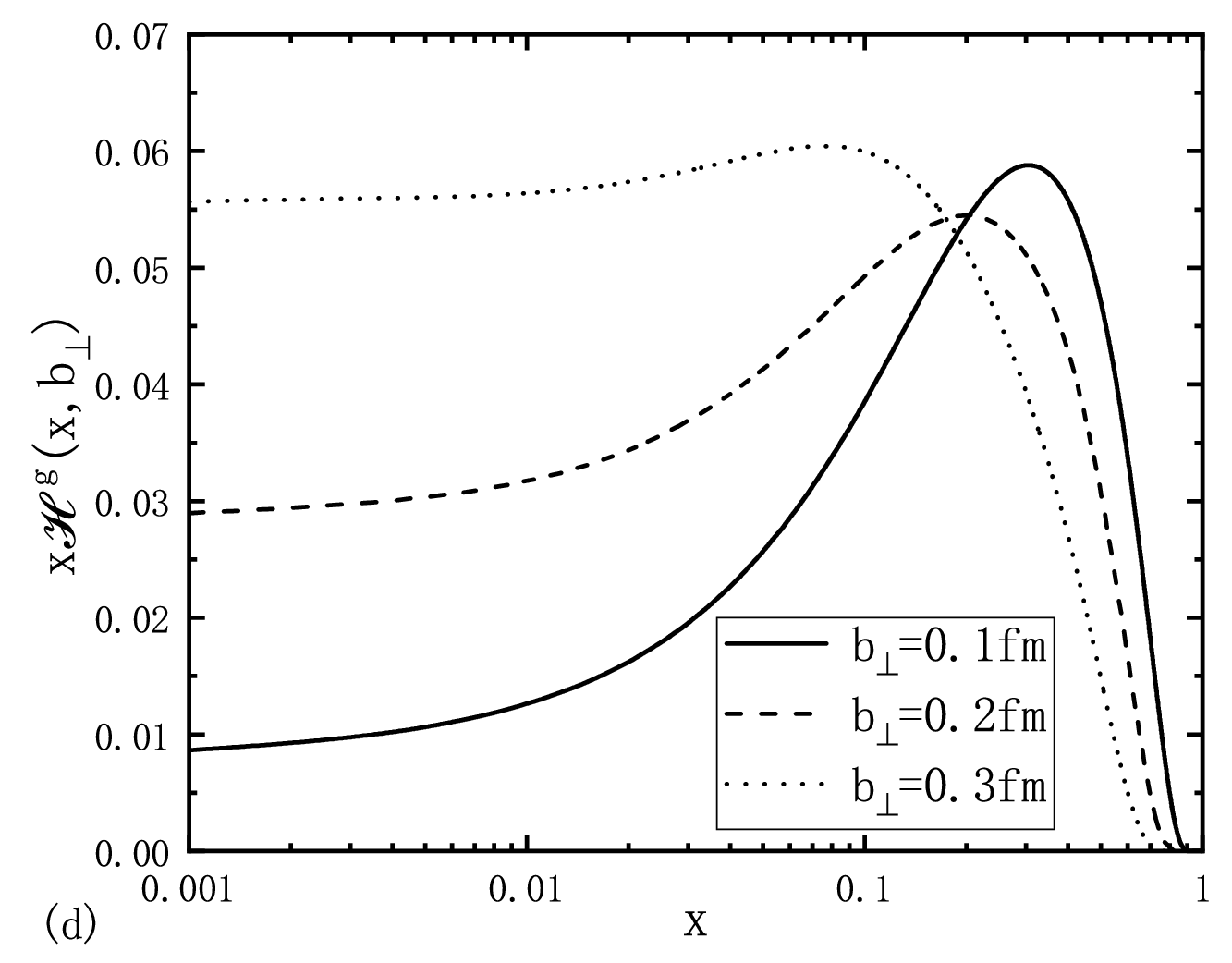}\\
	\includegraphics[width=0.42\columnwidth]{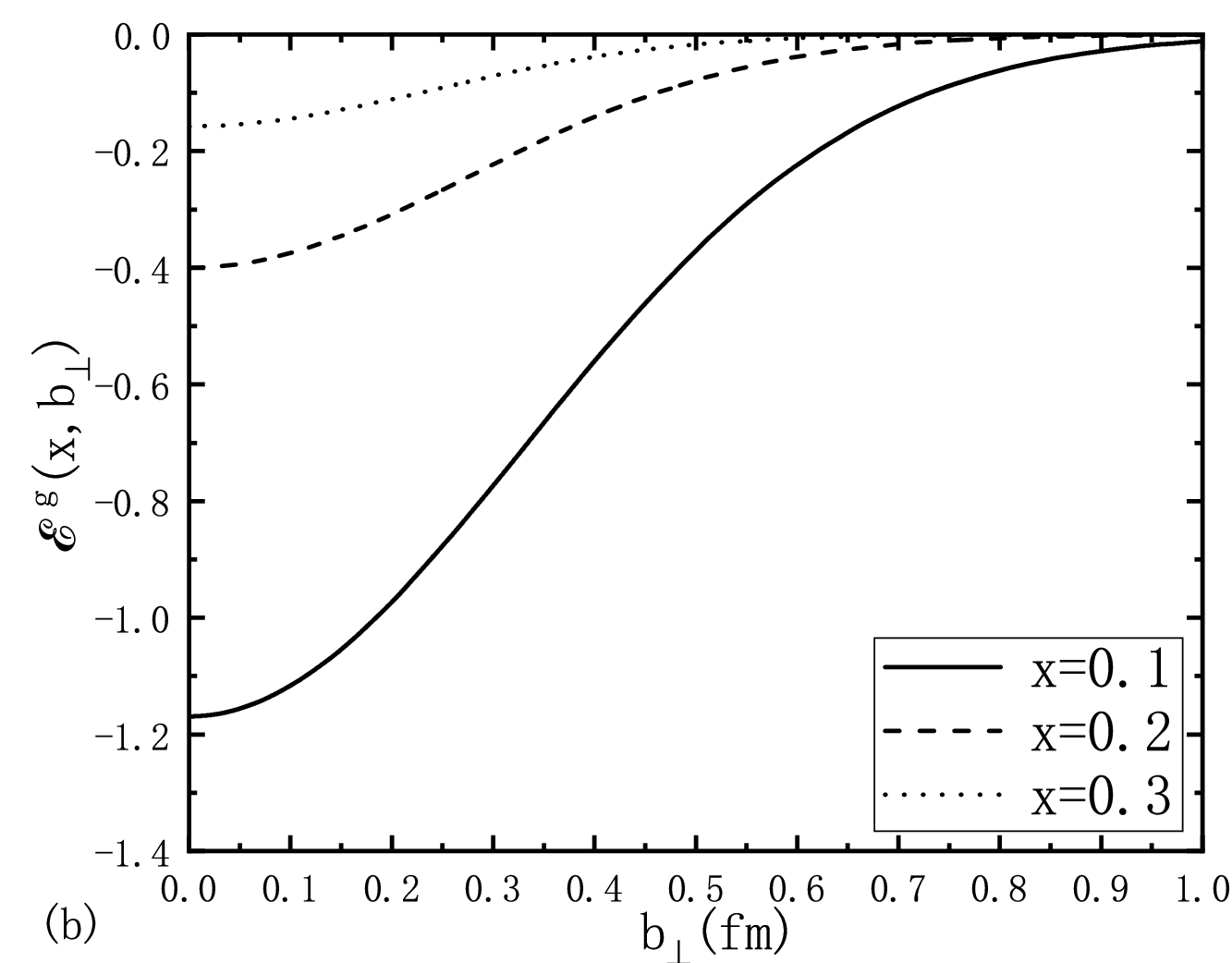}~~~
	\includegraphics[width=0.42\columnwidth]{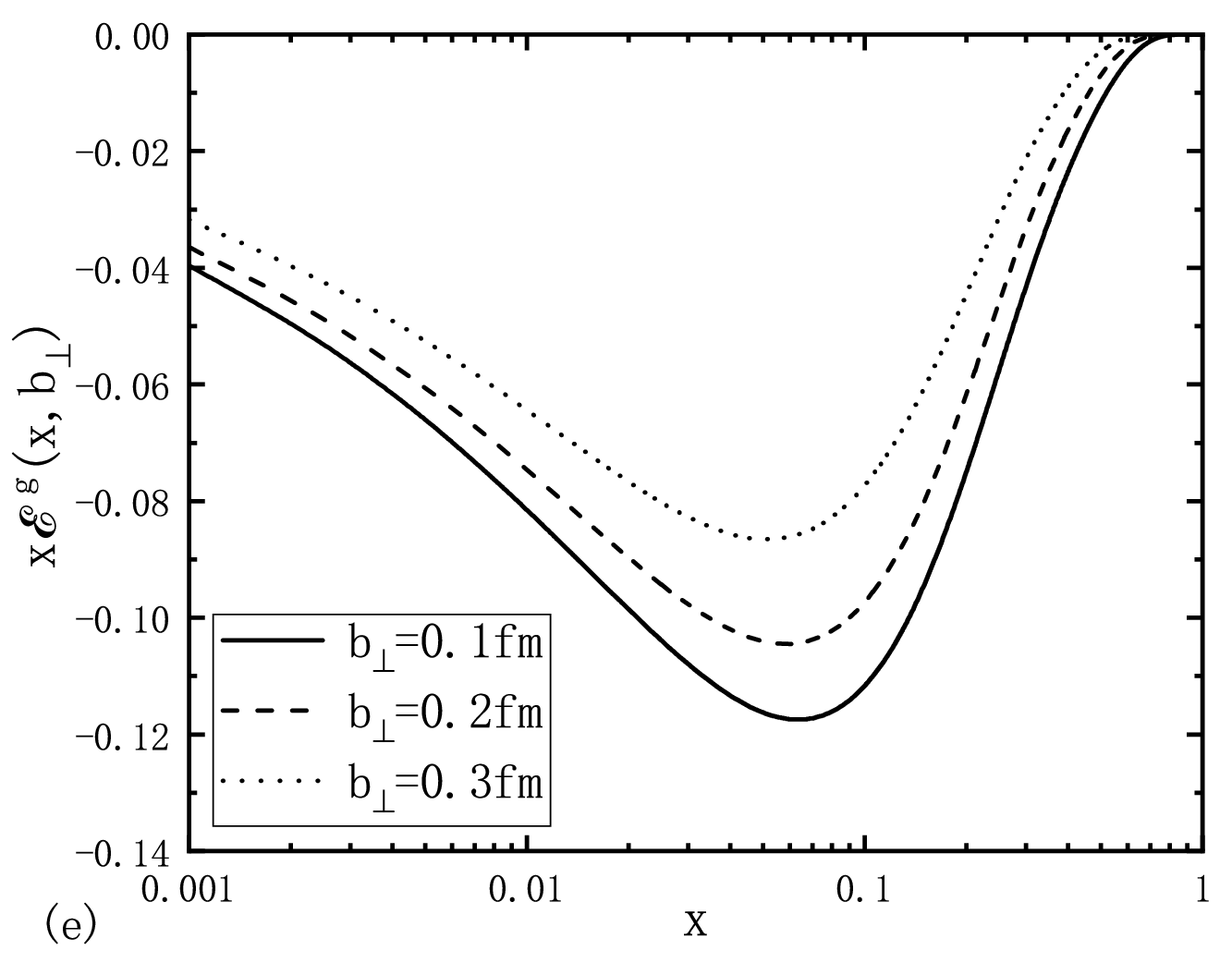}\\
	\includegraphics[width=0.42\columnwidth]{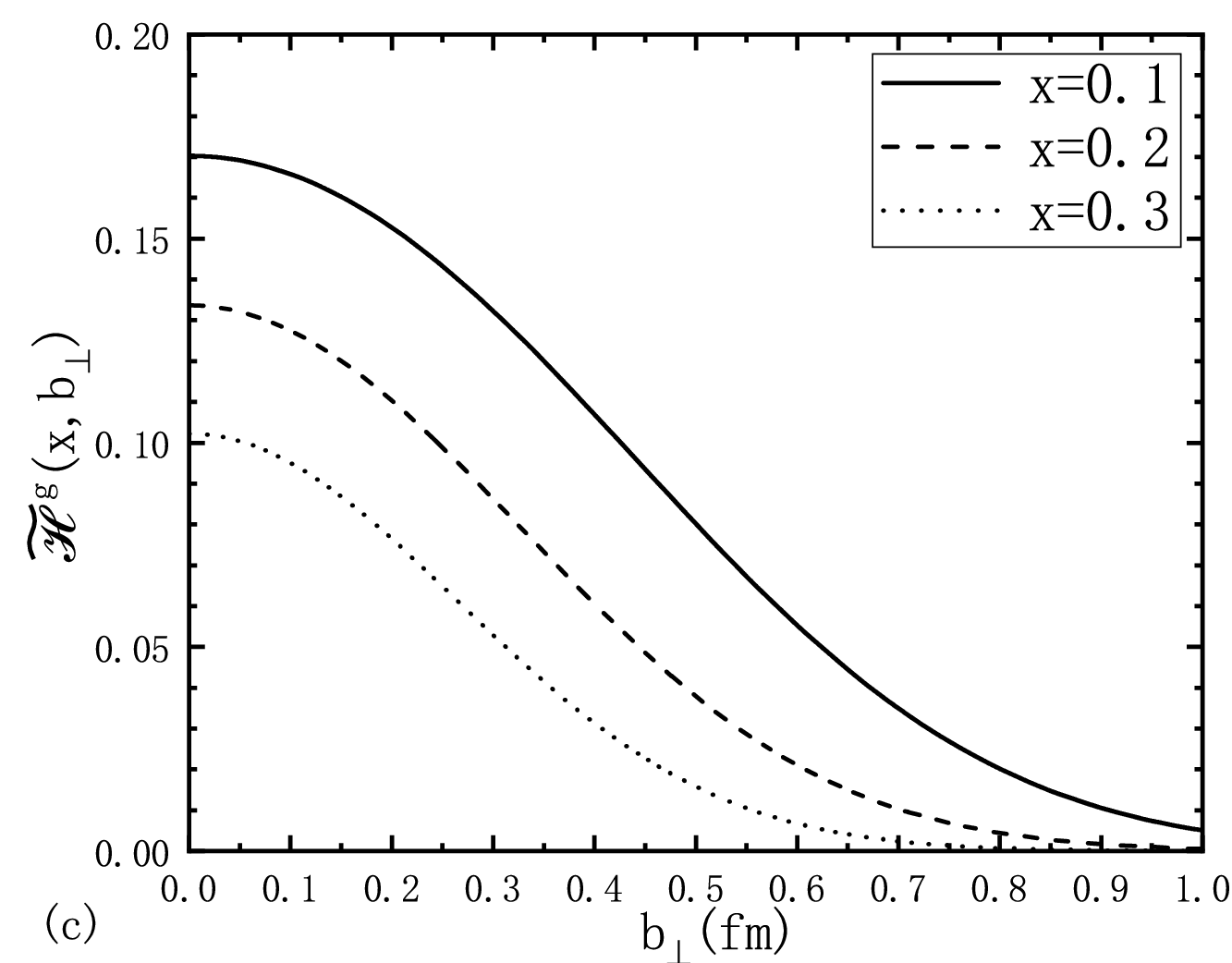}~~~
	\includegraphics[width=0.42\columnwidth]{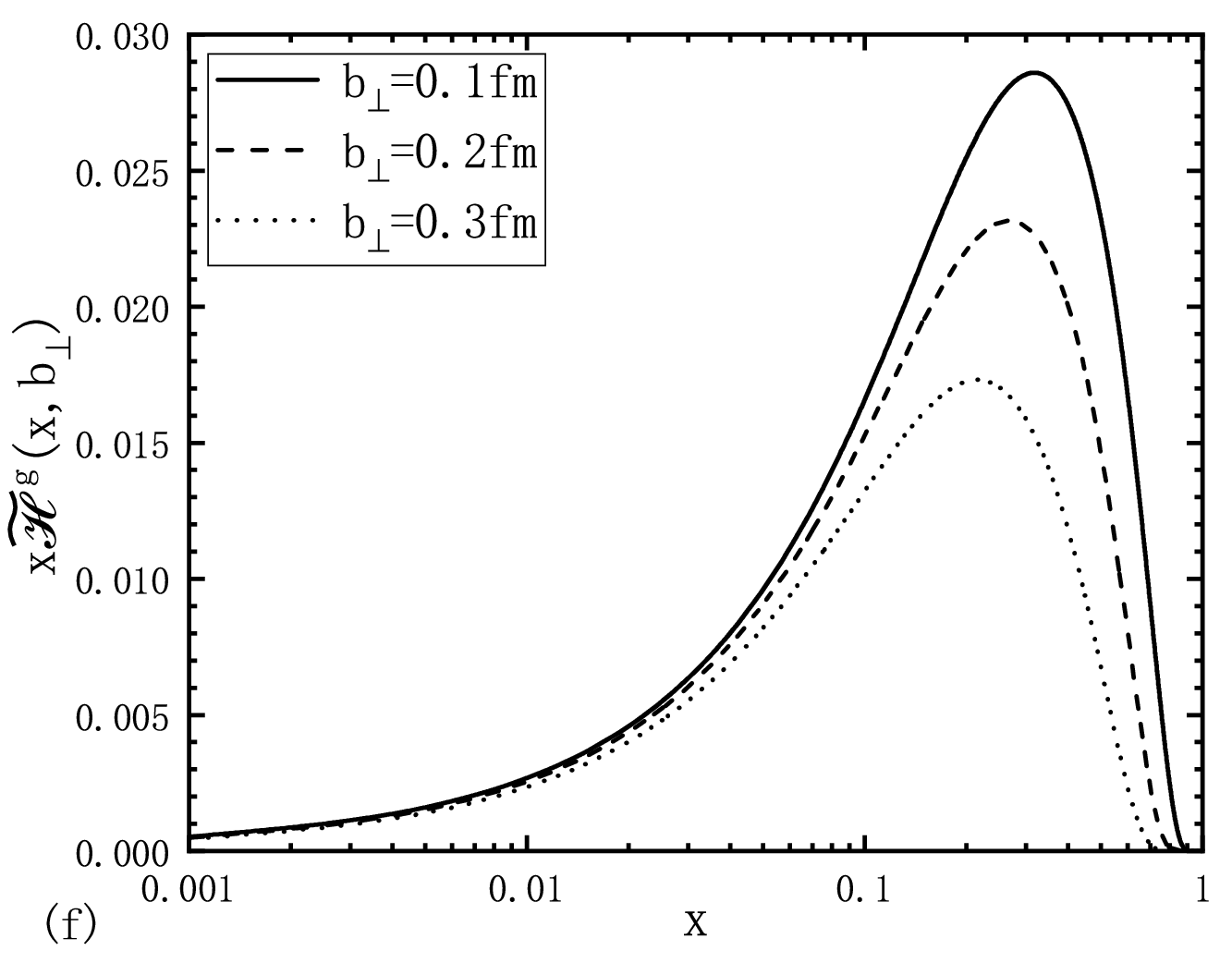}\\
	\caption{The left plane: The gluon IPDs $\mathcal{H}^g(x,\bm{b}_\perp)$, $\mathcal{E}^g(x,\bm{b}_\perp)$ and $\widetilde{\mathcal{H}}^g(x,\bm{b}_\perp)$ as functions of positive $b_\perp$ for  $x=0.1$, 0.2 and 0.3. The right plane: The gluon IPDs (timed with $x$) $x\mathcal{H}^g(x,\bm{b}_\perp)$, $x\mathcal{E}^g(x,\bm{b}_\perp)$ and $x\widetilde{\mathcal{H}}^g(x,\bm{b}_\perp)$ as functions of $x$ for $b_\perp=0.1$, 0.2 and 0.3 fm.}
	\label{fig:IPDs}
\end{figure}

In Fig.~\ref{fig:IPDs}(a-c), we present the model results of the gluon IPDs $\mathcal{H}^g(x,\bm{b}_\perp)$, $\mathcal{E}^g(x,\bm{b}_\perp)$ and $\widetilde{\mathcal{H}}^g(x,\bm{b}_\perp)$ as functions of  $b_\perp$ for $x=0.1$, 0.2 and 0.3, respectively. We notice that these distributions have the same trends. On the one hand, they approach to 0 with the increase of $b_\perp$. On the other hand, the magnitudes decrease with increasing $x$. $\mathcal{E}^g$ shows the negative distribution, while $\widetilde{\mathcal{H}}^g$ is positive and it decreases monotonically. 
Furthermore, $\mathcal{H}^g$ is also positive and it has a peak at $b_\perp>0$ when $x<0.2$.

Fig.~\ref{fig:IPDs}(d-f) show $x\mathcal{H}^g(x,\bm{b}_\perp)$, $x\mathcal{E}^g(x,\bm{b}_\perp)$ and $x\widetilde{\mathcal{H}}^g(x,\bm{b}_\perp)$ as functions of $x$ for $b_\perp=0.1$, 0.2 and 0.3 fm, respectively. The overall shape of all the plots is similar, and the polarity of each distribution is the same as that of the corresponding distribution in Fig.~\ref{fig:IPDs}(a-c). 
We observe that all the distributions vanish at $x=1$ and also have the tendency to approach to 0 at lower-$x$($<0.001$). 
The peaks of these distributions appear at low-$x$ and shift toward the lower values of $x$ with the increase of $b_\perp$. 
The magnitudes of $x\mathcal{E}^g$ and $x\widetilde{\mathcal{H}}^g$ decrease as $b_\perp$ increases, while $x\mathcal{H}^g$ has this trend only when $x>0.2$, which is consistent with the feature shown in Fig.~\ref{fig:IPDs}(a).

\section{Conclusions}\label{Sec:5}

We investigated the leading-twist GTMDs of the unpolarized and longitudinally polarized gluons inside the nucleon by applying a light-front gluon-triquark model for the nucleon motivated by soft-wall AdS/QCD. 
We obtained the numerical results for the GTMDs, showing that all the distributions are accessible in the DGLAP region $x>\xi$. 
The peaks of these distributions appear at the limit of $x \rightarrow \xi$ and shift toward higher values of $x$ with the increase of $\Delta^+$, and the magnitudes decrease as $x$ increases. 
An observation is that the main contributions of the gluon GTMDs are concentrated at low-$x$. 
In addition, we found that the gluon OAM and the nucleon spin tend to be antialigned in our model, so is the gluon spin and OAM. The former also implies that the total gluon OAM may reduce the total angular momentum contribution of the gluon to the nucleon
spin. Finally,
we studied the gluon IPDs  $\mathcal{H}^g(x,\bm{b}_\perp)$, $\mathcal{E}^g(x,\bm{b}_\perp)$ and $\widetilde{\mathcal{H}}^g(x,\bm{b}_\perp)$. 
The model results showed that $\mathcal{H}^g$ and $\widetilde{\mathcal{H}}^g$ are positive, while $\mathcal{E}^g$ is negative. 
The peaks of the IPDs (timed with $x$) appear at low-$x$ and shift toward lower values of $x$ with the increase of $b_\perp$, and the magnitudes of the IPDs decrease with the increase of $x$.
This study can provide better understanding on the information of nucleon structure in transverse momentum/position space in terms of gluon.

\section*{Acknowledgements}
This work is partially supported by the National Natural Science Foundation of China under grant number 12150013.


\begin{thebibliography}{99}

	\bibitem{Collins:1981uw}
	J.~C.~Collins and D.~E.~Soper,
	Nucl. Phys. B \textbf{194}, 445-492 (1982)
	
	\bibitem{Martin:1998sq}
	A.~D.~Martin, R.~G.~Roberts, W.~J.~Stirling and R.~S.~Thorne,
	Eur. Phys. J. C \textbf{4}, 463-496 (1998)
	[arXiv:hep-ph/9803445 [hep-ph]].
	
	\bibitem{Gluck:1994uf}
	M.~Gluck, E.~Reya and A.~Vogt,
	Z. Phys. C \textbf{67}, 433-448 (1995)
	
	\bibitem{Gluck:1998xa}
	M.~Gl\"uck, E.~Reya and A.~Vogt,
	Eur. Phys. J. C \textbf{5}, 461-470 (1998)
	[arXiv:hep-ph/9806404 [hep-ph]].
	
	\bibitem{Mulders:2000sh}
	P.~J.~Mulders and J.~Rodrigues,
	Phys. Rev. D \textbf{63}, 094021 (2001)
	[arXiv:hep-ph/0009343 [hep-ph]].
	
	\bibitem{Meissner:2007rx}
	S.~Meissner, A.~Metz and K.~Goeke,
	Phys. Rev. D \textbf{76}, 034002 (2007)
	[arXiv:hep-ph/0703176 [hep-ph]].
	
	\bibitem{Mulders:1995dh}
	P.~J.~Mulders and R.~D.~Tangerman,
	Nucl. Phys. B \textbf{461}, 197-237 (1996)
	[erratum: Nucl. Phys. B \textbf{484}, 538-540 (1997)]
	[arXiv:hep-ph/9510301 [hep-ph]].
	
	\bibitem{Bacchetta:2006tn}
	A.~Bacchetta, M.~Diehl, K.~Goeke, A.~Metz, P.~J.~Mulders and M.~Schlegel,
	JHEP \textbf{02}, 093 (2007)
	[arXiv:hep-ph/0611265 [hep-ph]].
	
	\bibitem{Barone:2001sp}
	V.~Barone, A.~Drago and P.~G.~Ratcliffe,
	Phys. Rept. \textbf{359}, 1-168 (2002)
	[arXiv:hep-ph/0104283 [hep-ph]].
	
	\bibitem{Brodsky:2002cx}
	S.~J.~Brodsky, D.~S.~Hwang and I.~Schmidt,
	Phys. Lett. B \textbf{530}, 99-107 (2002)
	[arXiv:hep-ph/0201296 [hep-ph]].
	
	\bibitem{Bacchetta:2017gcc}
	A.~Bacchetta, F.~Delcarro, C.~Pisano, M.~Radici and A.~Signori,
	JHEP \textbf{06}, 081 (2017)
	[erratum: JHEP \textbf{06}, 051 (2019)]
	[arXiv:1703.10157 [hep-ph]].
	
			\bibitem{Muller:1994ses}
		D.~M\"uller, D.~Robaschik, B.~Geyer, F.~M.~Dittes and J.~Ho\v{r}ej\v{s}i,
		Fortsch. Phys. \textbf{42} (1994), 101-141
		[arXiv:hep-ph/9812448 [hep-ph]].
	\bibitem{Ji:1996nm}
	X.~D.~Ji,
	Phys. Rev. D \textbf{55}, 7114-7125 (1997)
	[arXiv:hep-ph/9609381 [hep-ph]].
	
	\bibitem{Radyushkin:1997ki}
  A.~V.~Radyushkin,
  Phys.\ Rev.\  D {\bf 56}, 5524 (1997)
  [arXiv:hep-ph/9704207].

\bibitem{Mueller:1998fv}
  D.~Mueller, D.~Robaschik, B.~Geyer, F.~M.~Dittes and J.~Horejsi,
  Fortsch.\ Phys.\  {\bf 42}, 101 (1994)
  [arXiv:hep-ph/9812448].
	\bibitem{Goeke:2001tz}
	K.~Goeke, M.~V.~Polyakov and M.~Vanderhaeghen,
	Prog. Part. Nucl. Phys. \textbf{47}, 401-515 (2001)
	[arXiv:hep-ph/0106012 [hep-ph]].
	
	\bibitem{Diehl:2003ny}
	M.~Diehl,
	Phys. Rept. \textbf{388}, 41-277 (2003)
	[arXiv:hep-ph/0307382 [hep-ph]].
	
\bibitem{Ji:2004gf}
		X.~Ji,
		Ann. Rev. Nucl. Part. Sci. \textbf{54} (2004), 413-450
		doi:10.1146/annurev.nucl.54.070103.181302
	\bibitem{Belitsky:2005qn}
	A.~V.~Belitsky and A.~V.~Radyushkin,
	Phys. Rept. \textbf{418}, 1-387 (2005)
	[arXiv:hep-ph/0504030 [hep-ph]].
	
	\bibitem{Boffi:2007yc}
		S.~Boffi and B.~Pasquini,
		Riv. Nuovo Cim. \textbf{30} (2007) no.9, 387-448
		doi:10.1393/ncr/i2007-10025-7
		[arXiv:0711.2625 [hep-ph]].
	\bibitem{Burkardt:2000za}
	M.~Burkardt,
	Phys. Rev. D \textbf{62}, 071503 (2000)
	[erratum: Phys. Rev. D \textbf{66}, 119903 (2002)]
	[arXiv:hep-ph/0005108 [hep-ph]].
	
	\bibitem{Burkardt:2002hr}
	M.~Burkardt,
	Int. J. Mod. Phys. A \textbf{18}, 173-208 (2003)
	[arXiv:hep-ph/0207047 [hep-ph]].
	
	\bibitem{Meissner:2008ay}
	S.~Meissner, A.~Metz, M.~Schlegel and K.~Goeke,
	JHEP \textbf{08}, 038 (2008)
	[arXiv:0805.3165 [hep-ph]].
	
	\bibitem{Meissner:2009ww}
	S.~Meissner, A.~Metz and M.~Schlegel,
	JHEP \textbf{08}, 056 (2009)
	[arXiv:0906.5323 [hep-ph]].
	
	\bibitem{Bhattacharya:2017bvs}
	S.~Bhattacharya, A.~Metz and J.~Zhou,
	Phys. Lett. B \textbf{771}, 396-400 (2017)
	[erratum: Phys. Lett. B \textbf{810}, 135866 (2020)]
	[arXiv:1702.04387 [hep-ph]].
	
	\bibitem{Hagiwara:2017fye}
	Y.~Hagiwara, Y.~Hatta, R.~Pasechnik, M.~Tasevsky and O.~Teryaev,
	Phys. Rev. D \textbf{96}, no.3, 034009 (2017)
	[arXiv:1706.01765 [hep-ph]].
	
	\bibitem{Ji:2016jgn}
	X.~Ji, F.~Yuan and Y.~Zhao,
	Phys. Rev. Lett. \textbf{118}, no.19, 192004 (2017)
	[arXiv:1612.02438 [hep-ph]].
	
	\bibitem{Hatta:2016aoc}
	Y.~Hatta, Y.~Nakagawa, F.~Yuan, Y.~Zhao and B.~Xiao,
	Phys. Rev. D \textbf{95}, no.11, 114032 (2017)
	[arXiv:1612.02445 [hep-ph]].
	
	\bibitem{Bhattacharya:2022vvo}
	S.~Bhattacharya, R.~Boussarie and Y.~Hatta,
	Phys. Rev. Lett. \textbf{128}, no.18, 182002 (2022)
	[arXiv:2201.08709 [hep-ph]].
	
	\bibitem{Lorce:2011kd}
	C.~Lorce and B.~Pasquini,
	Phys. Rev. D \textbf{84}, 014015 (2011)
	[arXiv:1106.0139 [hep-ph]].
	
	\bibitem{Hatta:2011ku}
	Y.~Hatta,
	Phys. Lett. B \textbf{708}, 186-190 (2012)
	[arXiv:1111.3547 [hep-ph]].
	
	\bibitem{Ji:2012sj}
	X.~Ji, X.~Xiong and F.~Yuan,
	Phys. Rev. Lett. \textbf{109}, 152005 (2012)
	[arXiv:1202.2843 [hep-ph]].
	
	\bibitem{Lorce:2012ce}
	C.~Lorce,
	Phys. Lett. B \textbf{719}, 185-190 (2013)
	[arXiv:1210.2581 [hep-ph]].
	
	\bibitem{Lorce:2014mxa}
	C.~Lorc\'e,
	Phys. Lett. B \textbf{735}, 344-348 (2014)
	[arXiv:1401.7784 [hep-ph]].
	
	\bibitem{Tan:2021osk}
	C.~Tan and Z.~Lu,
	Phys. Rev. D \textbf{105}, no.3, 034004 (2022)
	[arXiv:2110.08502 [hep-ph]].
	\bibitem{Lorce:2011ni}
	C.~Lorce, B.~Pasquini, X.~Xiong and F.~Yuan,
	Phys. Rev. D \textbf{85}, 114006 (2012)
	[arXiv:1111.4827 [hep-ph]].
	
	\bibitem{Lorce:2011dv}
	C.~Lorce, B.~Pasquini and M.~Vanderhaeghen,
	JHEP \textbf{05}, 041 (2011)
	[arXiv:1102.4704 [hep-ph]].
	
	\bibitem{Mukherjee:2014nya}
	A.~Mukherjee, S.~Nair and V.~K.~Ojha,
	Phys. Rev. D \textbf{90}, no.1, 014024 (2014)
	[arXiv:1403.6233 [hep-ph]].
	
	\bibitem{Mukherjee:2015aja}
	A.~Mukherjee, S.~Nair and V.~K.~Ojha,
	Phys. Rev. D \textbf{91}, no.5, 054018 (2015)
	[arXiv:1501.03728 [hep-ph]].
	
	\bibitem{More:2017zqq}
	J.~More, A.~Mukherjee and S.~Nair,
	Phys. Rev. D \textbf{95}, no.7, 074039 (2017)
	[arXiv:1701.00339 [hep-ph]].
	
	\bibitem{Liu:2015eqa}
	T.~Liu and B.~Q.~Ma,
	Phys. Rev. D \textbf{91}, 034019 (2015)
	[arXiv:1501.07690 [hep-ph]].
	
	\bibitem{Chakrabarti:2016yuw}
	D.~Chakrabarti, T.~Maji, C.~Mondal and A.~Mukherjee,
	Eur. Phys. J. C \textbf{76}, no.7, 409 (2016)
	[arXiv:1601.03217 [hep-ph]].
	
	\bibitem{Chakrabarti:2017teq}
	D.~Chakrabarti, T.~Maji, C.~Mondal and A.~Mukherjee,
	Phys. Rev. D \textbf{95}, no.7, 074028 (2017)
	[arXiv:1701.08551 [hep-ph]].
	
	\bibitem{Chakrabarti:2019wjx}
	D.~Chakrabarti, N.~Kumar, T.~Maji and A.~Mukherjee,
	Eur. Phys. J. Plus \textbf{135}, no.6, 496 (2020)
	[arXiv:1902.07051 [hep-ph]].
	
	\bibitem{Gutsche:2016gcd}
	T.~Gutsche, V.~E.~Lyubovitskij and I.~Schmidt,
	Eur. Phys. J. C \textbf{77}, no.2, 86 (2017)
	[arXiv:1610.03526 [hep-ph]].
	
	\bibitem{Kaur:2019lox}
	S.~Kaur and H.~Dahiya,
	Adv. High Energy Phys. \textbf{2020}, 9429631 (2020)
	[arXiv:1906.04662 [hep-ph]].
	
	\bibitem{Kumar:2017xcm}
	N.~Kumar and C.~Mondal,
	Nucl. Phys. B \textbf{931}, 226-249 (2018)
	[arXiv:1705.03183 [hep-ph]].
	
	\bibitem{Kanazawa:2014nha}
	K.~Kanazawa, C.~Lorc\'e, A.~Metz, B.~Pasquini and M.~Schlegel,
	Phys. Rev. D \textbf{90}, no.1, 014028 (2014)
	[arXiv:1403.5226 [hep-ph]].
	
	\bibitem{Ma:2018ysi}
	Z.~L.~Ma and Z.~Lu,
	Phys. Rev. D \textbf{98}, no.5, 054024 (2018)
	[arXiv:1808.00140 [hep-ph]].
	
	\bibitem{Kaur:2019jow}
	S.~Kaur and H.~Dahiya,
	Phys. Rev. D \textbf{100}, no.7, 074008 (2019)
	[arXiv:1908.01939 [hep-ph]].
	
	\bibitem{Kaur:2019kpi}
	N.~Kaur and H.~Dahiya,
	Eur. Phys. J. A \textbf{56}, no.6, 172 (2020)
	[arXiv:1909.10146 [hep-ph]].

	\bibitem{Zhang:2021tnr}
	J.~L.~Zhang and J.~L.~Ping,
	Eur. Phys. J. C \textbf{81}, no.9, 814 (2021)
	
	\bibitem{Broniowski:2003rp}
	W.~Broniowski and E.~Ruiz Arriola,
	Phys. Lett. B \textbf{574}, 57-64 (2003)
	[arXiv:hep-ph/0307198 [hep-ph]].
	
	\bibitem{Lyubovitskij:2021qza}
	V.~E.~Lyubovitskij and I.~Schmidt,
	Phys. Rev. D \textbf{104}, no.1, 014001 (2021)
	[arXiv:2105.07842 [hep-ph]].
	
	\bibitem{Lyubovitskij:2020xqj}
	V.~E.~Lyubovitskij and I.~Schmidt,
	Phys. Rev. D \textbf{103}, no.9, 094017 (2021)
	[arXiv:2012.01334 [hep-ph]].
	
	\bibitem{Brodsky:2014yha}
	S.~J.~Brodsky, G.~F.~de Teramond, H.~G.~Dosch and J.~Erlich,
	Phys. Rept. \textbf{584}, 1-105 (2015)
	[arXiv:1407.8131 [hep-ph]].
	
	\bibitem{Gutsche:2019jzh}
	T.~Gutsche, V.~E.~Lyubovitskij and I.~Schmidt,
	Nucl. Phys. B \textbf{952}, 114934 (2020)
	[arXiv:1906.08641 [hep-ph]].
	
	\bibitem{Brodsky:2007hb}
	S.~J.~Brodsky and G.~F.~de Teramond,
	Phys. Rev. D \textbf{77}, 056007 (2008)
	[arXiv:0707.3859 [hep-ph]].
	
	\bibitem{Abidin:2009hr}
	Z.~Abidin and C.~E.~Carlson,
	Phys. Rev. D \textbf{79}, 115003 (2009)
	[arXiv:0903.4818 [hep-ph]].
	
	\bibitem{Brodsky:1973kr}
	S.~J.~Brodsky and G.~R.~Farrar,
	Phys. Rev. Lett. \textbf{31}, 1153-1156 (1973)
	
	\bibitem{Lu:2016vqu}
	Z.~Lu and B.~Q.~Ma,
	Phys. Rev. D \textbf{94}, no.9, 094022 (2016)
	[arXiv:1611.00125 [hep-ph]].
	
	\bibitem{Lorce:2013pza}
	C.~Lorc\'e and B.~Pasquini,
	JHEP \textbf{09}, 138 (2013)
	[arXiv:1307.4497 [hep-ph]].
	
		\bibitem{Maji:2022tog}
	T.~Maji, C.~Mondal and D.~Kang,
	Phys. Rev. D \textbf{105}, no.7, 074024 (2022)
	[arXiv:2202.08635 [hep-ph]].
	
	\bibitem{Bacchetta:2008af}
	A.~Bacchetta, F.~Conti and M.~Radici,
	Phys. Rev. D \textbf{78}, 074010 (2008)
	[arXiv:0807.0323 [hep-ph]].
	
	\bibitem{Brodsky:2000xy}
	S.~J.~Brodsky, M.~Diehl and D.~S.~Hwang,
	Nucl. Phys. B \textbf{596}, 99-124 (2001)
	[arXiv:hep-ph/0009254 [hep-ph]].
	
	\bibitem{Gutsche:2011vb}
	T.~Gutsche, V.~E.~Lyubovitskij, I.~Schmidt and A.~Vega,
	Phys. Rev. D \textbf{85}, 076003 (2012)
	[arXiv:1108.0346 [hep-ph]].
	
	\bibitem{Lyubovitskij:2020otz}
	V.~E.~Lyubovitskij and I.~Schmidt,
	Phys. Rev. D \textbf{102}, no.3, 034011 (2020)
	[arXiv:2005.10163 [hep-ph]].
	
	\bibitem{Sufian:2020wcv}
	R.~S.~Sufian, T.~Liu and A.~Paul,
	Phys. Rev. D \textbf{103}, no.3, 036007 (2021)
	[arXiv:2012.01532 [hep-ph]].
	
	\bibitem{Brodsky:1989db}
	S.~J.~Brodsky and I.~A.~Schmidt,
	Phys. Lett. B \textbf{234}, 144-150 (1990)
	
	\bibitem{Brodsky:1994kg}
	S.~J.~Brodsky, M.~Burkardt and I.~Schmidt,
	Nucl. Phys. B \textbf{441}, 197-214 (1995)
	[arXiv:hep-ph/9401328 [hep-ph]]. 
	
	\bibitem{NNPDF:2017mvq}
	R.~D.~Ball \textit{et al.} [NNPDF],
	Eur. Phys. J. C \textbf{77}, no.10, 663 (2017)
	[arXiv:1706.00428 [hep-ph]].
	
	\bibitem{Nocera:2014gqa}
	E.~R.~Nocera \textit{et al.} [NNPDF],
	Nucl. Phys. B \textbf{887}, 276-308 (2014)
	[arXiv:1406.5539 [hep-ph]].
	
\bibitem{Tan:2023vvi}
C.~Tan and Z.~Lu,
[arXiv:2312.07997 [hep-ph]].
	
	\bibitem{Bhattacharya:2018lgm}
	S.~Bhattacharya, A.~Metz, V.~K.~Ojha, J.~Y.~Tsai and J.~Zhou,
	Phys. Lett. B \textbf{833}, 137383 (2022)
	[arXiv:1802.10550 [hep-ph]].
	
	
\end{thebibliography}
\end{document}